\documentclass[floatsintext,man,dvipsnames,table]{apa7}
\usepackage{lipsum}
\usepackage{amssymb}
\usepackage{authblk}
\usepackage[american]{babel}
\usepackage{multirow}
\usepackage{authblk}
\usepackage{booktabs}
\usepackage{csquotes}
\usepackage{subcaption} %  for subfigures environments 
\usepackage{graphicx} 
\usepackage{caption}
\usepackage{tabularray}
\usepackage[style=apa,sortcites=true,sorting=nyt,backend=biber]{biblatex}
\DeclareLanguageMapping{american}{american-apa}
\title{Hyperauthored papers disproportionately amplify important egocentric network metrics}
\shorttitle{Hyperauthorship and egocentric network metrics}

\abstract{ %limit to 200 words 
Hyperauthorship, a phenomenon whereby there are a disproportionately large number of authors on a single paper, is increasingly common in several scientific disciplines, but with unknown consequences for network metrics used to study scientific collaboration. The validity of co-authorship as a proxy for scientific collaboration is affected by this.
Using bibliometric data from publications in the field of genomics, we examine the impact of hyperauthorship on metrics of scientific collaboration, and propose a method to determine a suitable cutoff threshold for hyperauthored papers and compare co-authorship networks with and without hyperauthored works. Our analysis reveals that including hyperauthored papers dramatically impacts the structural positioning of central authors and the topological characteristics of the network, while producing small influences on whole-network cohesion measures. We present two solutions to minimize the impact of hyperauthorship: using a mathematically grounded and reproducible calculation of threshold cutoff to exclude hyperauthored papers or fractional counting to weight network results. Our findings affirm the structural influences of hyperauthored papers and suggest that scholars should be mindful when using co-authorship networks to study scientific collaboration.
}

\authornote{
  Correspondence concerning this article should be addressed to Ly Dinh, School of Information, University of South Florida.  E-mail: lydinh@usf.edu}

\begin{document}
%\maketitle

\section{Hyperauthored papers disproportionately amplify important egocentric network metrics}
Scientific collaboration is vital to solving complex scientific problems that require integration of knowledge across disciplines. Bibliometric studies show that discipline-spanning collaborations play an important role in spurring scientific innovation and producing impactful papers \parencites{thelwall2022research}{uzzi2013atypical}{collins2015expertise}. To that end, increasing efforts have been made to support scientific research teams as well as to better understand the relationship between diversity in scientific collaborations and research outcomes. Many studies in this space use paper co-authorship as the primary indicator upon which to assess the diversity of a collaboration and that collaboration's effects on outcomes of scholarly interest. 

%This paper reminds scholars to exhibit caution when analyzing and interpreting co-authorship networks to measure collaboration in light of the increasing prominence of hyperauthored papers (i.e. a disproportionately large number of authors for a single paper) in science.

The average size of authorship teams has increased over time \parencites{wuchty2007increasing}{ioannidis2008measuring}, especially in fields such as high-energy physics \parencites{birnholtz2006does}{milojevic2010modes}, genomics \parencite{dinh2018middle}, and medicine \parencite{franceschet2010effect}, where hyperauthorship is common. The rapid growth in average team size may impact measures of scientific collaboration outcomes, which traditionally have been examined using a mix of bibliometric \parencites{rafols2012journal}{porter2009science}{schummer2004multidisciplinarity} and network analysis methods \parencites{akbaritabar2021quantitative}{barley2022exploring}{cummings2003structural}{fegley2013has}. In fact, \cite{sinatra2015century} found that the number of citations per paper and number of papers per author in the field of interdisciplinary physics have been inflated over the past 15 years, and that the number of authors per paper increased at similar rate as the number of papers produced in this field. Thus, these citation measures may not be the most reliable proxy for scientific collaboration \parencite{strumia2019biblioranking}. In some leading medical journals, for example, hyperauthored works can include long lists of authors that represent an honorary role in the research process despite not having contributed substantively to the work \parencites{kennedy2014honorary}{wislar2011honorary}. Furthermore, the validity co-authorship as a primary indicator of research collaborations is another subject of inquiry as evidence suggests that not all collaborations result in co-authored papers \parencites{lundberg69g}{smith2000collab}{tijssen2004commercialisation} and that not all co-authorships signify collaboration in terms of contribution to writing \parencites{cronin2001hyperauthorship}{dinh2018middle}. It is important to distinguish that scientific collaboration is a process of working together, whereas co-authorship is an indicator of scientific contribution with certain norms and guidelines \parencite{cronin2001hyperauthorship}. Thus, they refer to different aspects of scientific research, and while co-authorship may suggest collaboration, there may be other factors beyond direct collaboration that can explain co-authorship \parencite{birnholtz2006does}. 

Given the complex relationship between co-authorship and scientific collaboration, especially in the presence of hyperauthorship, this study examines how hyperauthored papers impact the co-authorship network metrics that scholars use to study scientific collaborations. The inclusion of even a few hyperauthored papers within a bibliometrically constructed network may substantially inflate the average number of collaborators per author. Consequently, including hyperauthored works may inflate author-level network measures frequently used to assess an author's influence in scientific collaboration. We test this hypothesis by examining a database of papers from the interdisciplinary field of genomics, in which hyperauthorship is common. Using these data, we (1) propose a method to determine a suitable cutoff threshold for hyperauthored papers using cumulative frequency distribution of number of authors per paper; (2) compare the changes (if any) in network metrics of co-authorship networks with and without hyperauthored papers; and (3) present two solutions to minimize the impact of hyperauthorship by using a threshold cutoff to exclude hyperauthored papers or using fractional counting (i.e. Newman and Jaccard weighing functions) to weight network results. 

Our analysis reveals that including hyperauthored papers dramatically impacts the structural positioning of central authors and the topological characteristics of the network, while producing comparatively small influences on whole-network cohesion measures. These findings suggest scholars should be mindful when using bibliometric networks to study scientific collaboration, especially if the object of analysis focuses on ego-centric dependent variables. We argue that researchers should consider whether including hyperauthored works is necessary to address their research questions, and consider omitting them from analysis when unnecessary. Further, when including hyperauthored work, we find that a fractional counting approach overall can mitigate the impact of hyperauthorship compared to full counting, with the most optimal solution being fractional counting based on the number of shared co-authors across all papers. Our findings affirm researchers' concerns about the structural influences of hyperauthored papers and indicate that scholars must directly consider how hyperauthored works will affect their results when studying scientific collaboration using co-authorship networks.

%Specifically, these studies all found that while co-authorship is a reliable indicator of scientific output, it does not capture other types of institutional collaborations such as industry-academic collaborations \cite{lundberg69g} and corporate partnerships \cite{tijssen2004commercialisation}. Furthermore, \cite{glanzel2004does} found that papers with larger number of authors were more likely to be cited than solo-authored papers. Similarly, \cite{batista2006possible} found that comparing research outputs across disciplines is difficult as some fields such as Theoretical and High Energy Physics have notably larger number of co-authors. \cite{milojevic2010modes} also observed that the probability of new collaboration between a given pair of authors is inflated when papers with more than 200 authors (called ``hyperauthored'' papers) are included in the sample. Given the complex relationship between co-authorship and other variables of scientific outputs such as likelihood of future collaborations and citation counts, we aim to investigate how co-authorship influence measures of collaborations such as diversity of teams...  \cite{perianes2016constructing} compares the impact of full counting and fractional counting on co-authorship networks .. in the case of the full counting method, a publication co-authored by three researchers is assigned to each researcher with a full weight of one. On the other hand, in the case of the fractional counting method, the publication is assigned to each researcher with a fractional weight of 1/3.

\section{Background}
\subsection{Network metrics as indicators of scientific collaboration}
Scholars in the science of science have used network measures to analyze structural patterns of scientific collaboration \parencites{bordons2015relationship}{leydesdorff2007betweenness} as well as to identify factors that impact collaboration across disciplinary \parencites{morillo2003interdisciplinarity}{porter2007measuring} and geographical \parencites{naik23qss}{bordons2000collaboration} boundaries. The benefits of collaboration in the production of scientific knowledge are well-defined in the literature, including that more diverse research teams can benefit from increased creativity and innovation \parencites{burt2004structural}{nemeth2003better}. \cite{leydesdorff2007betweenness} found that betweenness centrality is a reliable indicator of interdisciplinary in journal–journal citation networks; the higher the betweenness, the more diverse the disciplines that cite a journal. \cite{bordons2015relationship}'s network analysis of co-authors in three fields (Nanoscience, Pharmacology, and Statistics) showed that authors with the most number of 'strong tie' co-authors (i.e., those with repeated collaborations) tend to have the highest research productivity. \cite{costanza2012authorship} examined the co-authorship network of 172 authors who published the most number of papers in an interdisciplinary field (Ecosystem Services) and found that (1) number of co-authors had a positive linear relationship with the number of citations an article received, which also had a positive correlation with the average h-index of each article. These studies exemplify that network analysis is a preferred method of analysis in which co-authorship and citation patterns often are used as proxies for scientific collaboration. 

%\cite{dehdarirad2017research} examines assortativity patterns in co-authorship network

%\cite{velden2010new} examine co-authorship network patterns at multiple levels of analysis; global, meso, and local; they also used similar metrics to us. 

%\cite{ahmed2013toward} has used centrality measures to examine co-authorship structures.

\subsection{Hyperauthorship and scientific collaboration}
Bibliometric studies have found a continuous and consistent growth in co-authorship that spans across all scientific disciplines \parencites{milojevic2010modes}{dehdarirad2017research}{valderas2007cited}. While there are notable benefits of scientific collaboration, there are also drawbacks to consider, particularly in terms of fair allocation of credit when co-authorship is given for reasons other than scientific collaboration \parencites{cronin2001hyperauthorship}{birnholtz2006does}. Especially with the rising prevalence of publications with large numbers of co-authors, known as hyperauthorship, norms and requirements for authorship in a collaborative work are also impacted \parencite{cronin2001hyperauthorship}. Scholars in bibliometrics and network science have found that hyperauthorship affects traditional indicators of scholarship productivity such as H-index \parencite{koltun2021h}, degree centrality \parencite{fegley2013has}, and author degree distributions \parencite{milojevic2010modes}. \cite{koltun2021h}'s analysis of over two million publications on Google Scholar and the citations between them revealed that authors with 100 co-authors or more over the course of their careers have disproportionately high H-indices. However, the H-indices were found to be uncorrelated with other productivity indicators, such as the number of scientific awards received. \cite{fegley2013has} found that hyperauthorship influenced co-authorship network structure, where groups of authors were completely connected within their own clusters (i.e. common multi-authored paper) and thus had higher degree centrality than expected. \cite{milojevic2010modes} compared the probability distributions of new collaboration based on prior co-authorship with and without hyperauthorship and showed that while both distributions are power-law, the distribution with hyperauthorship includes anomalous noise. %Specifically, random peaks were present when the number of previous collaborations were at seven and nine prior co-authored papers, which were not explainable. 
The author also found that for authors with less than a total of twenty co-authors over the course of their careers, the degree distribution was a log-normal ``hook'' instead of a power law. This finding illustrates that the number of co-authors has an effect on the topology of the collaboration network. Altogether, these studies show that hyperauthorship may impact a scholar's interpretation of a particular author's (or a group of authors') collaboration activity and their connectedness within a network. 

\subsection{Mitigating the impact of hyperauthorship}

We have observed that in many studies using bibliometric data, hyperauthored papers are not explicitly acknowledged or addressed. In some cases, keeping hyperauthored papers may be useful or necessary, such as in studies examining author name disambiguation \parencites{farber2022microsoft}{kim2019scale}, researcher productivity \parencites{thelwall2022research}{costas2010bibliometric}, or growth in co-authorship size over time \parencites{thelwall2022research}{borner2005studying}. Among those works that do explicitly seek to mitigate the impact of hyperauthorship the most common practice has been to exclude papers that have a certain number of co-authors \parencites{cronin2001hyperauthorship}{milojevic2010modes}{fegley2013has}{morris2007manifestation}. But, practices for choosing this threshold have been inconsistent. \cite{cronin2001hyperauthorship} was the first to define hyperauthorship as any paper with more than 100 authors. Similarly, \cite{milojevic2010modes} set a threshold of more than 200 authors for a hyperauthored paper. Both \cite{fegley2013has} and \cite{morris2007manifestation} operationalized a hyperauthored paper to have at least 20 authors. \cite{ahmed2013toward} removed 3\% of papers that were identified as hyperauthored, but did not specify the threshold. While these empirical solutions are important first steps, variability in how hyperauthored papers are handled creates challenges for comparability across studies. Therefore, a standardized and reproducible method for identifying and excluding hyperauthorship would be beneficial. To address this need, we demonstrate how a reproducible pipeline for preprocessing hyperauthorship data can help mitigate any potential effects of hyperauthorship on network measures of interest, while also offering researchers a standard to consider when seeking to exclude hyperauthored works from their datasets.

\section{Methodology}
\subsection{Data}
The dataset used for this study consisted of bibliometric records for 413 researchers within a large biological research institute. Each researcher's publication data throughout their academic career (up until 2021) were collected using a Scopus database, including metadata such as: full title, publication type, journal/conference proceeding name, publisher, DOI, author names, organizational unit of author(s), citation counts (based on Scopus), open access status, and keywords. We added a unique ID to each publication for quick retrieval and matching purposes. The resulting publication dataset contained 19,100 unique papers produced by 35,658 unique authors. The original format of the dataset was a two-mode network as two types of nodes, papers and authors, are connected. An edge between a paper node and an author node indicates that a paper is authored by a particular author. As our goal was to analyze co-authorship network patterns, we transformed the two-mode network into an author-author one-mode network via weighted bipartite projection method by \cite{borgatti2011analyzing}. The resulting weighted projected graph contains edges between two author nodes that are previously connected to the same paper node in the original bipartite graph. In other words, the projected graph contains co-authorship edges, along with weights reflecting the number of papers that two authors have co-authored together.

\subsubsection{Threshold for Hyperauthorship}
We establish a threshold to determine when a paper is categorized as hyperauthored based on the distribution of the number of authors per paper using cumulative frequency distribution approach. Our goal is to show a generalized and reproducible method for identifying and removing hyperauthored papers from any skewed distribution of publications. Our process involves several steps to determine the cut-off point for hyperauthorship, where outliers with a large number of authors could be excluded from analysis. First, we check whether the distribution of the number of authors per paper is normally distributed. If the distribution is normal, we use the empirical rule (i.e. 68-95-99.7 rule) to determine outliers by identifying the threshold at which 95\% of the data are captured (i.e. two standard deviations from the mean). If the distribution is not normally distributed, we apply Chebyshev's inequality (i.e. 75-88.9 rule) to determine a threshold at which 88.9\% of the data are captured (i.e. three standard deviations from the mean). We then compare our approach to the cumulative frequency distribution method as another point of comparison. The cumulative frequency approach involves ranking observations in order of magnitude and calculating cumulative frequencies based on the ranking. We opt to use two methods for determining the cut-off point so that we can cross-validate the findings and establish the reliability of our approach. 

\subsection{Network weighing functions}
Another potential solution to addressing hyperauthored works is to apply weighting functions to potentially reduce these products' structural influence on collaboration networks. Fair allocation of authorship credit to authors engaged in multi-authored papers has been a topic of considerable interest for bibliometrics researchers \parencites{perianes2016constructing}{abramo2013importance}{sivertsen2019measuring}. This problem is especially relevant to researchers who use a combination of bibliometric and network approaches as choices about credit allocation has direct impact on how the network is constructed and weighted \parencites{perianes2016constructing}{gauffriau2017categorization}. \cite{gauffriau2021counting} in their comprehensive literature review find that (1) full counting and (2) fractional counting are two primary methods used for credit allocation. The full counting method assigns a weight of one to each author of a paper, whereas the fractional counting method distributes a single weight among all the co-authors of a paper. In this study, we will implement both counting methods, with formulations stated below: 

%There's also harmonic counting based on ranking of author \parencite{hagen2008harmonic} , where first author would get most credit, then harmonically divided to second, third .... This is good, but we don't want to make assumption about the ranks in this paper... 

\subsubsection{Full counting}
\vspace{-1cm}
\begin{equation}
\label{eqn:full_counting}
w_{ij} = \sum a_{i}^{p}a_{j}^{p}
\end{equation}
Where $a_{i}^{p}$ indicates whether $i$ is an author in paper \textit{p}, where one indicates authorship and zero indicates no authorship. Similarly, $a_{j}^{p}$ is one if $j$ is an author in paper \textit{p}. Thus, the resulting $w_{ij}$ is one if $i$ and $j$ are both authors in \textit{p}. This counting method attributes a full weight of one to each co-authorship instance and aggregates based on the number of papers on which $i$ and $j$ are both co-authors.
%Been found that full counting is frequently used, but inflates bibliometric measures as the full credit is repeatedly given to all co-authors \parencite{hagen2008harmonic}

\subsubsection{Fractional counting}
There are several approaches to fractional counting \parencite{gauffriau2021counting}, which are essentially means of normalizing the co-authorship weights based on the number of co-authors in a paper. Based on prior literature, we utilize two main weighting functions, namely Newman's and Jaccard's methods. Newman's method and variants of the method have been used in prior studies such as \cite{griffin2021collaborations} and \cite{perianes2016constructing}. Jaccard's method has been used in \cite{brandao2017strength} and \cite{pan2012evolution} as a measure of neighborhood overlap between two authors. 

\textbf{Weight $w_{ij}$ based on \cite{newman2001scientific}}: 
\begin{equation}
\label{eqn:newman_counting}
w_{ij} = \sum p \frac{1}{N_p{-1}}
%\[ \scalebox{1.5}{$w_{ij} = \sum p \frac{1}{N_p{-1}}$} \]
\end{equation}
Where $N_p$ is the number of co-authors in paper \textit{p}, and this sum does not include single-authored papers \parencite{newman2001scientific}. 
 $N_p{-1}$ is for the removal of self-loops in each egonetwork, ensuring that each author has only $N-1$ co-authors. The numerator of $1$ is derived from Newman's operationalization of $a_{i}^{p}$$a_{j}^{p}$ (same as in equation \ref{eqn:full_counting}) where both $a_{i}^{p}$ and $a_{j}^{p}$ is 1 if author $i$ and author $j$ are co-authors of paper $p$ (\cite{newman2001scientific}, Formula 2). 

\textbf{Weight $w_{ij}$ based on Jaccard index \cite{borgatti2011analyzing}}: 
\begin{equation}
\label{eqn:jaccard_counting}
{w_{ij} = \frac{N(i) \cap N(j)}{N(i) \cup  N(j)}}
%}\[ \scalebox{1.5}{$w_{ij} = \frac{N(i) \cap N(j)}{N(i) \cup  N(j)}$} \]
\end{equation}

Where $N_i$ is the number of co-authors that $i$ has and $N_j$ is the number of co-authors that $j$ has. Thus, $N(i) \cap N(j)$ is the number of shared co-authors between $i$ and $j$; $N(i) \cup  N(j)$ is the number of all co-authors that both $i$ and $j$ have.

\subsection{Network measures}
Here, we define the network measures that are computed for this study. We use existing algorithms available in $NetworkX$, a $Python$ library for network analysis, and modify a subset of measures based on our operationalization. We conduct network analysis at both the whole-network and ego-centric network levels, computing the same set of metrics (as discussed below) to both levels. 

\subsubsection{Whole-network cohesion measures}

\textit{Density} measures the proportion of edges that exist in a network relative to the total number of possible edges. We use the formula $(2m) / (n(n-1))$ to calculate density, where $n$ is the number of nodes, and $m$ is the number of edges. 

\textit{Average clustering} measures the extent to which nodes in a network tend to form local neighborhoods. We calculate this by dividing the fraction of triangles in the network by the possible number of triangles that could exist with a given network size.

\textit{Average path length} measures the average shortest path distance between every pair of nodes. We use Dijkstra's algorithm for weighted network where each node is iteratively selected as a source node and a shortest path to every other node is calculated. 

\textit{Giant component} is the largest connected subgraph in a network, where all nodes are reachable to each other. Similar to how Dijkstra's algorithm works, this algorithm iterates through each source node and conducts a breadth-first search to ensure there is no disconnected path between any two nodes. We first extracted all the connected components that exist in the network, then determine the largest component and assign that as the giant component.

\subsubsection{Centrality measures}
\textit{Degree centrality} measures the amount of edges that each node has to other nodes in the network. We normalize each node's degree centrality by dividing it by the maximum degree of network ($n-1$). 

\textit{Betweenness centrality} measures the number of shortest paths that pass through each node. This measure indicates the extent to which a node can bridge connection(s) to other nodes in the network. Given the size of the network and the computational complexity, we approximate betweenness centrality based on a random sample of 1000 nodes. This measure is normalized by $1/((n-1)(n-2))$ where $n$ is the number of nodes in a directed network. 

\textit{Closeness centrality} measures the average reachability of one node to other nodes in the network. We calculate this based on the reciprocal of the average path length between a source node and all other $n-1$ nodes. 

\textit{Eigenvector centrality} measures the extent to which a node is an immediate neighbor of well-connected nodes. Eigvenctor centrality is calculated by ${Ax} = \lambda {x}$, where $A$ is an adjacency matrix of the network with an eigenvalue of $\lambda$. The algorithm iterates over each node and is complete when $\lambda$ is the highest in $A$. 
    
\subsubsection{Topological measures}
\textit{Omega} coefficient indicates an extent to which a network exhibits a small-world property. The formula is $\omega$ = $Lr/L - C/Cl$, where $C$ is the clustering coefficient of the network, $L$ is the average path length of the network, $Lr$ is the average path length of the simulated random network, and $Cl$ is the clustering coefficient of the simulated lattice network. $\omega$ coefficient ranges from $-\infty$ to positive $\infty$, where $\omega$ close to zero reflects a small-world topology. A random network is indicated by a positive $\omega$. A lattice network is indicated by a negative $\omega$. We compute $\omega$ on the giant component, with five rewiring iterations per edge, and five random graphs generated to calculate the simulated statistics.

\textit{Alpha} exponent indicates the extent to which the network's degree distribution exhibits a power-law fit. The algorithm is implemented via the $powerlaw$ package in $python$, where the optimal $\alpha$ exponent is computed for the network. $\alpha$ ranges from 1 to $\infty$, and $\alpha$ between 2 and 3 indicates that the network degree distribution is a power-law fit \parencite{newman2005power}.

\subsection{Comparison between networks without and with hyperauthorship}
We calculate the percentage change in network measures without and with the inclusion of hyperauthored papers. The formula used to calculate percentage change between two values is: 
\vspace{-2mm}
\begin{equation}
\label{eqn:percent_change}
Percent Change =\frac{(V_2-V_1)}{\left| V1 \right|} \times 100\
%\[ \scalebox{1.1}{$Percent Change =\frac{(V_2-V_1)}{\left| V1 \right|} \times 100$} \]
\end{equation}

\section{Results}
\begin{figure}[t]  
{\includegraphics[width=1.02\textwidth]{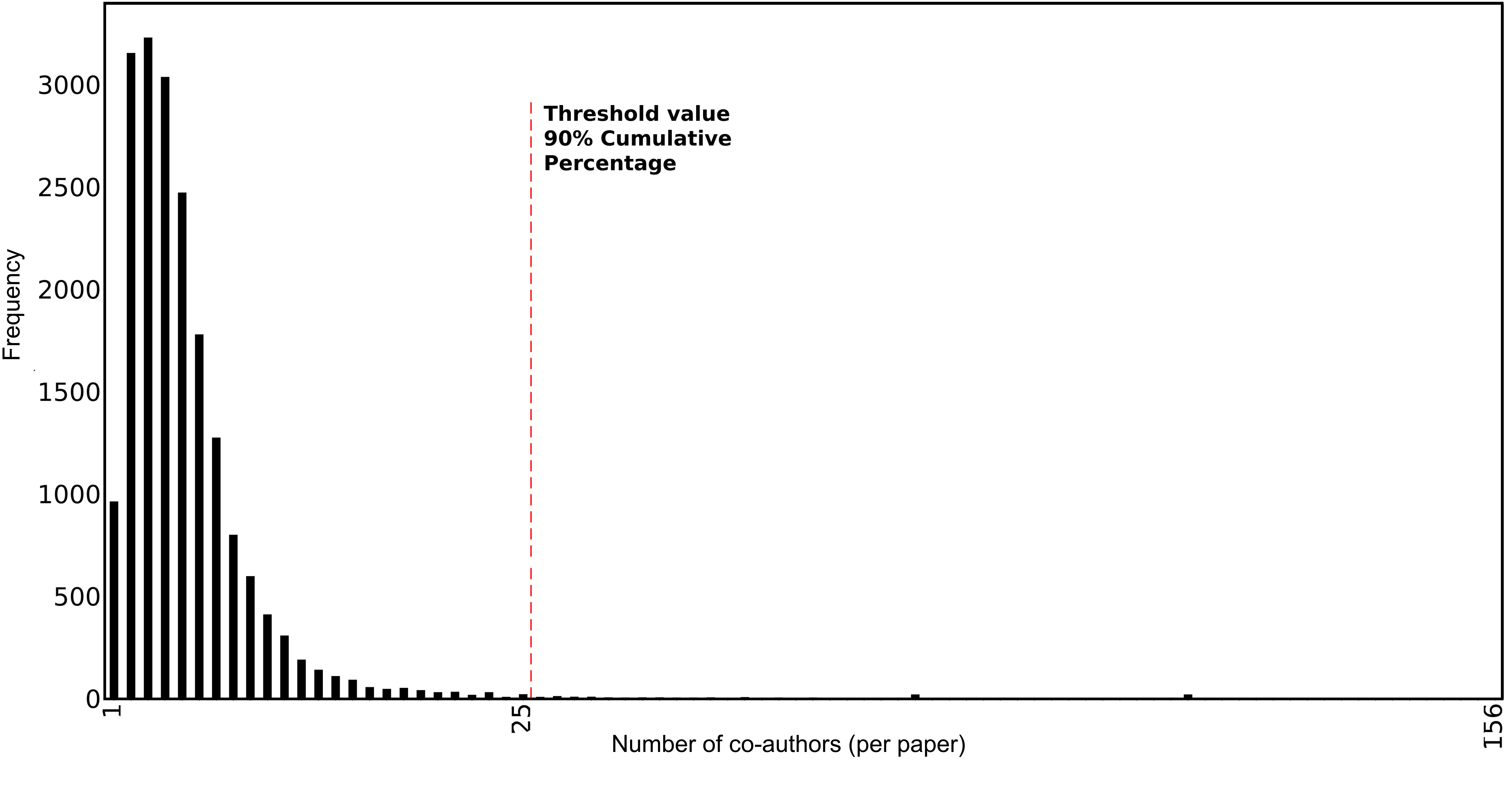}}
\caption{Histogram depicting number of co-authors per paper for articles included in this analysis. Red dotted line indicates the cut-off threshold in authorship at 90\% cumulative percentage indicating hyperauthorship}
\label{fig:hyperauthor_cutoff}
\end{figure}

\begin{figure}[htbp]
\centering  
{\includegraphics[width=\textwidth]{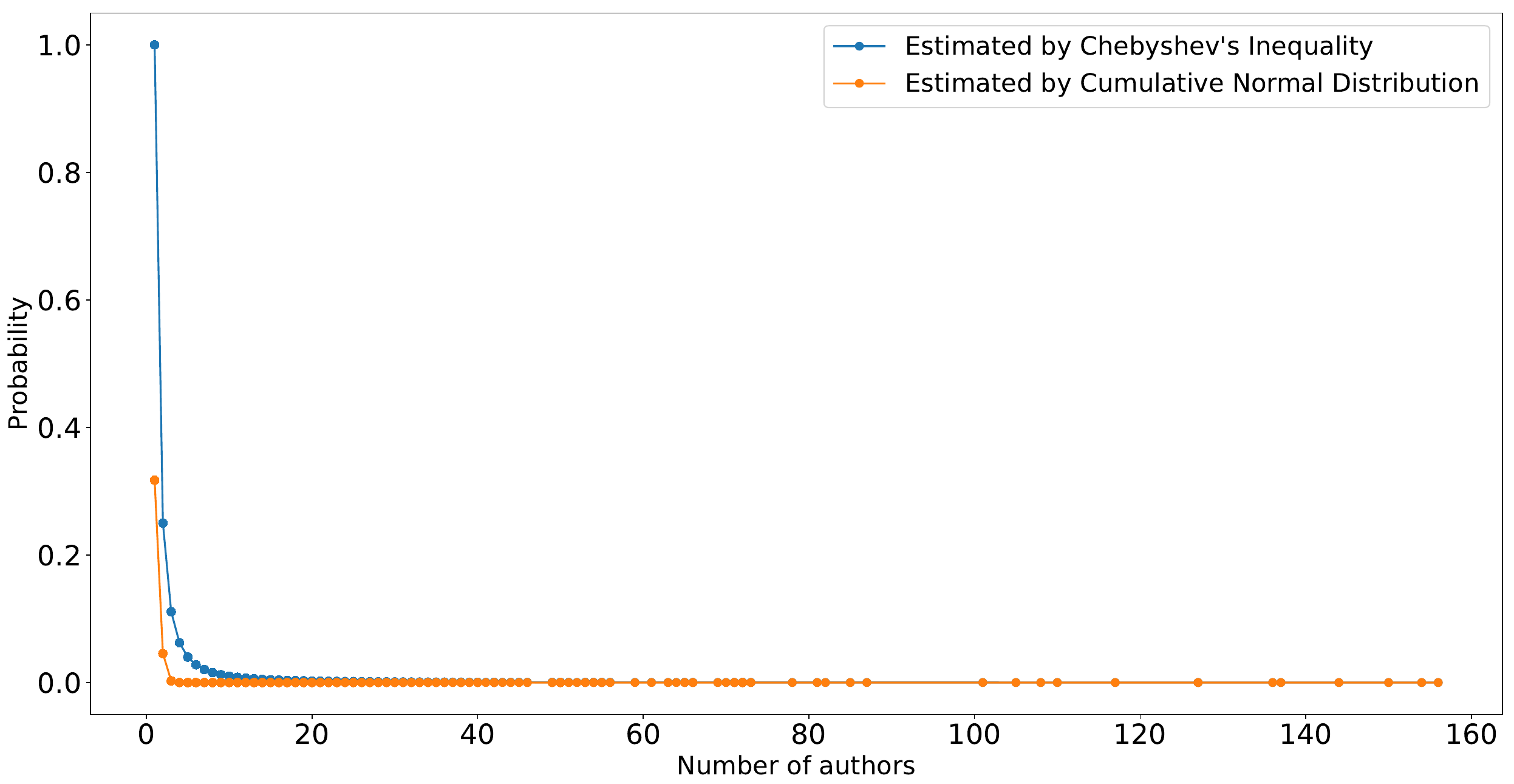}}
\caption{Histogram depicting the distribution of the number of authors per paper in this analysis, as estimated by (1) Chebyshev's inequality (blue line) and (2) cumulative normal distribution (orange line)}
\label{fig:hyperauthor_dist}
\end{figure}

We first present the hyperauthorship cut-off results based on our authorship threshold approach. The number of authors per paper ranges from 1 to 156 in our dataset (mean=5.46, median=4, SD=6.37; Figure 1). Given that mean number of authors is more than the median number of authors, we expect a positively-skewed distribution. As shown in Figure \ref{fig:hyperauthor_dist}, we have a skewed probability distribution and thus opted to use Chebyshev's inequality function to estimate a suitable outlier threshold. %We expect that hyperauthored papers are outliers in the distribution as they would have notably larger number of authors per paper compared to other papers in the dataset. 
Based on Chebyshev's distribution at $k=3$, where approximately 89\% of the data will be within three standard deviations, we find the upper bound at 25.85. This means that a threshold of $\approx$ 26 authors and above would be considered outliers in this dataset. We further evaluate the reliability of this threshold by using a cumulative percentage approach, as shown in \ref{fig:hyperauthor_cutoff}. We find that 90\% of papers are included within a threshold of 25 authors per paper. Thus, this method suggests a cutoff threshold of excluding all papers with more than 25 co-authors from analyses. Using this cutoff, we removed 203 papers that have more than 25 authors. These papers have a range of 26 to 156 authors per paper (mean=50.88, median=43, SD=26.25). After removing these works, the resulting distribution of the updated dataset changes to a range of 1 to 25 authors per paper (mean=4.97, median=4, SD=3.36). 

\begin{table}[htb]
\centering
\caption{Network descriptives for co-authorship networks \underline{without} and networks \underline{with} hyperauthorship}
\label{tab:network_descriptives}
\resizebox{\textwidth}{!}{%
\begin{tabular}{lrrr} 
\toprule
Network measures & \textbf{W/o}~hyperauthors & \textbf{With} hyperauthors & \% change \\ \hline
\# of Unique Papers & 18,897 & 19,100 & +1.074\% \\
\# of Nodes (authors) & 30,467 & 35,658 & +17.038\% \\
\# of Edges (co-authorship) & 199,581 & 441,892 & +121.41\%\\
Density & 0.0004 & 0.0007 & +75\% \\
Avg. Clustering & 0.854 & 0.854 & 0\%\\
Avg. Path Length (of subgraph) & 6.947 & 6.152 & +11.444\%\\
Size of Giant Component & 198,750  & 441,163 & +121.969\%\\
\# of Components & 15 & 14 & -6.667\%\\
\hline
Small-Worldliness ($\omega$) & -0.295 & -0.355 & -20.339\%\\
Power-law Exponent ($\alpha$) & 2.919 & 4.867 & +66.735\%\\
\hline
Avg. Degree Centrality (unweighted) & 13.101 & 24.785 & +89.184\% \\
Avg. Eigenvector Centrality (unweighted) & 0.003 & 0.0007 & -76.667\% \\
Avg. Betweenness Centrality (unweighted) & 0.0001 & 0.000 & -100\%\\
Avg. Closeness Centrality (unweighted) & 0.231 & 0.240 & +3.896\% \\
\bottomrule
\end{tabular}
}
\end{table}

The structural and topological characteristics of the networks are impacted to varying degrees as a result of excluding versus including hyperauthored papers (Table \ref{tab:network_descriptives}). The co-authorship network without hyperauthored papers was projected based on 18,897 unique papers. The network including hyperauthored papers was projected based on 19,100 papers, which contains 203 more papers than the first network. Although including hyperauthored papers results in a minimal percentage change in number of papers between the two networks (1.074\%), the resulting changes to network size are notable. There is a 17\% (n=5,191) increase in the number of authors when hyperauthored papers are included, which resulted in a notable increase of 121\% (n=242,311) in co-authorship ties. The density of the network also increases by 75\% given the rise in number of edges, however there is no change in average clustering across the two networks. Although the number of edges increases when hyperauthored papers are included, the number of closed triangles between nodes remains the same. This indicates that more edges do not necessarily lead to a higher level of triadic closure among the authors. There is a slight increase in the average shortest path length (+11\%) and a decrease in the number of components (from 15 to 14) in the network with hyperauthored papers. Interestingly, the size of the largest (giant) component increases with similar magnitude (+121\%) with the increase in number of edges. In particular, the giant component in the network without hyperauthorship excludes 831 edges, and the giant component in the network with hyperauthorship excludes 729 edges, thus suggesting that the network with hyperauthored papers has slightly fewer pendant edges that are not connected to the rest of the network.

In terms of topology, both networks without and with hyperauthored papers exhibit a lattice-like structure as opposed to a small-world structure (negative $\omega$ values). The network without hyperauthored papers exhibits a closer fit to a power-law topology (i.e. hubs-and-spokes structure, consistent with \cite{newman2005power}'s finding) than network with hyperauthored papers. This result also highlights the impact hyperauthorship has on the degree distribution that changes the topology of the network. 

\begin{figure}[htb]
\begin{subfigure}[h]{0.5\linewidth}
\includegraphics[width=\linewidth]{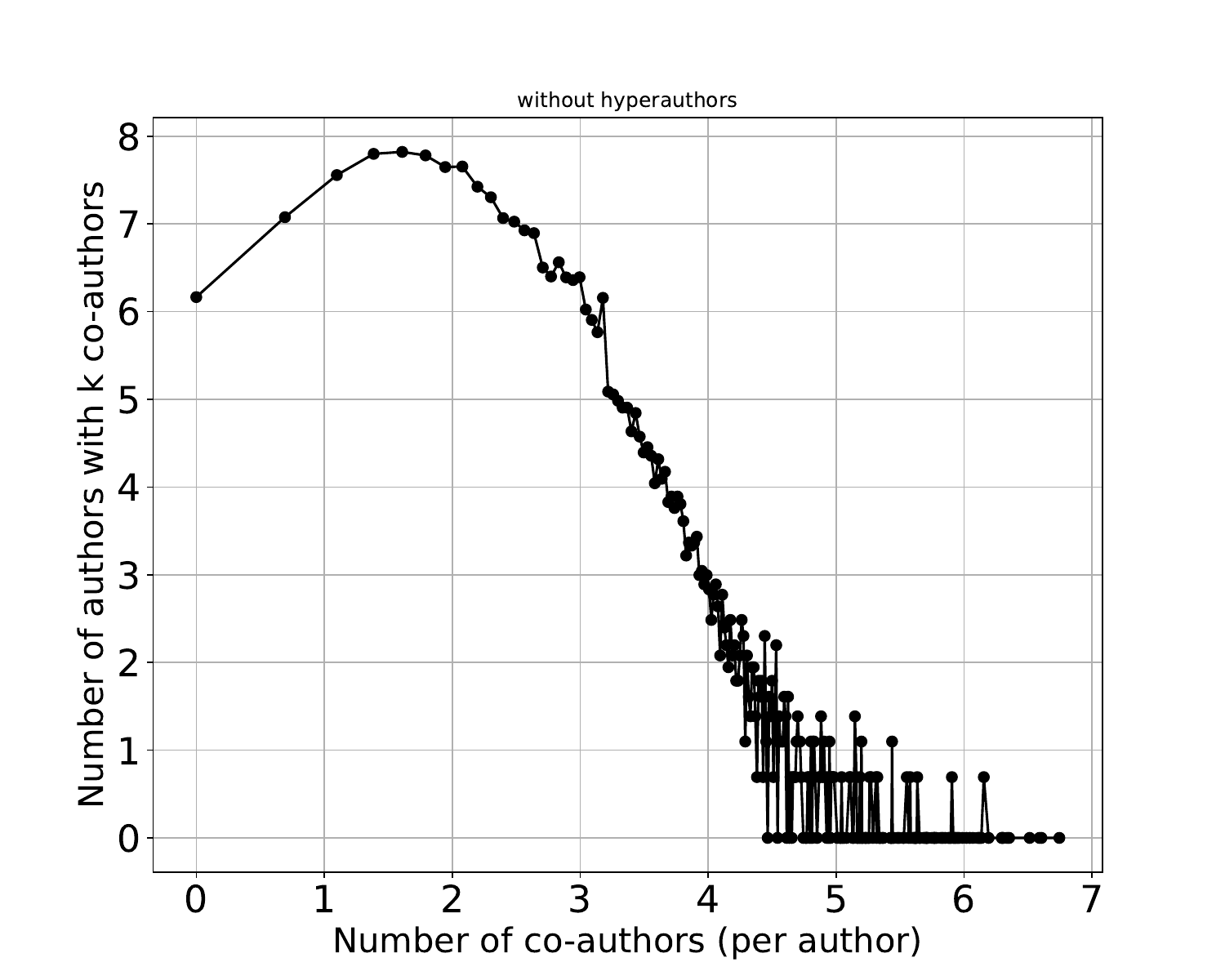}
\caption{Without hyperauthors}
\end{subfigure}
\hfill
\begin{subfigure}[h]{0.5\linewidth}
\includegraphics[width=\linewidth]{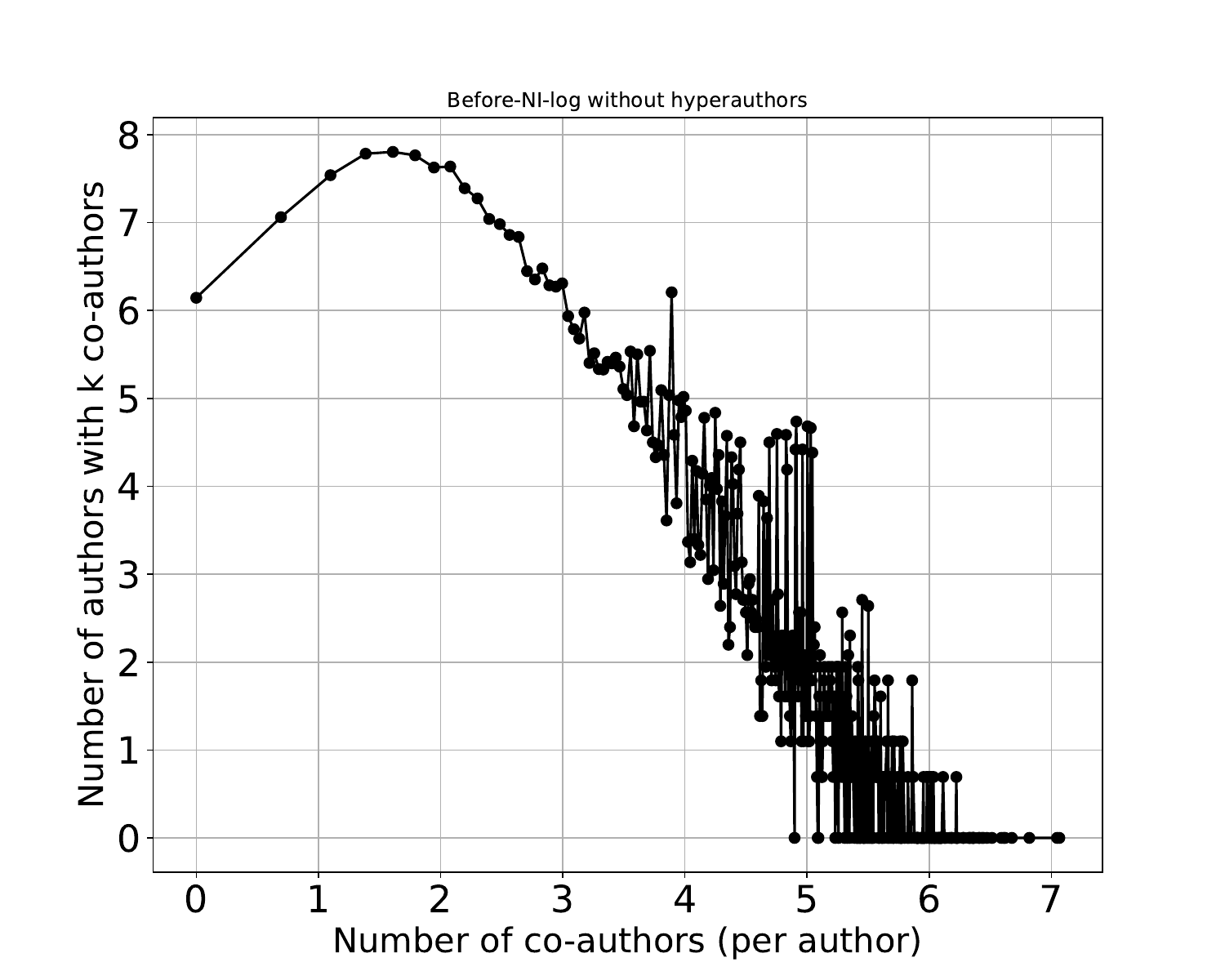}
\caption{With hyperauthors}
\end{subfigure}%
\caption{Log-Normal distribution of the number of co-authors of a given author}
\label{fig:coauthor_dist_log}
\vspace{-0.7cm}
\end{figure}

\begin{figure}[htb]
\begin{subfigure}[h]{0.5\linewidth}
\includegraphics[width=\linewidth]{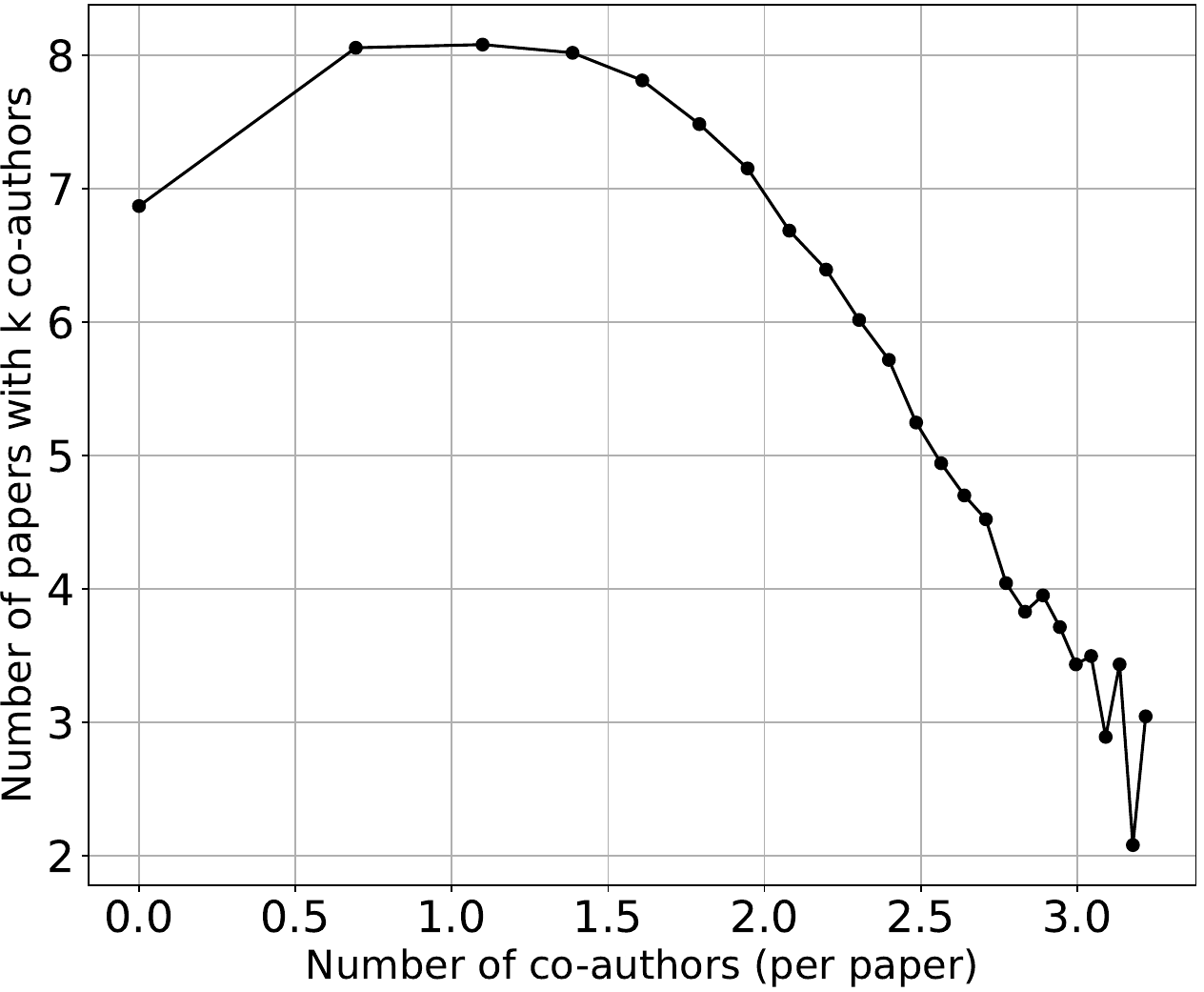}
\caption{Without hyperauthors}
\end{subfigure}
\hfill
\begin{subfigure}[h]{0.5\linewidth}
\includegraphics[width=\linewidth]{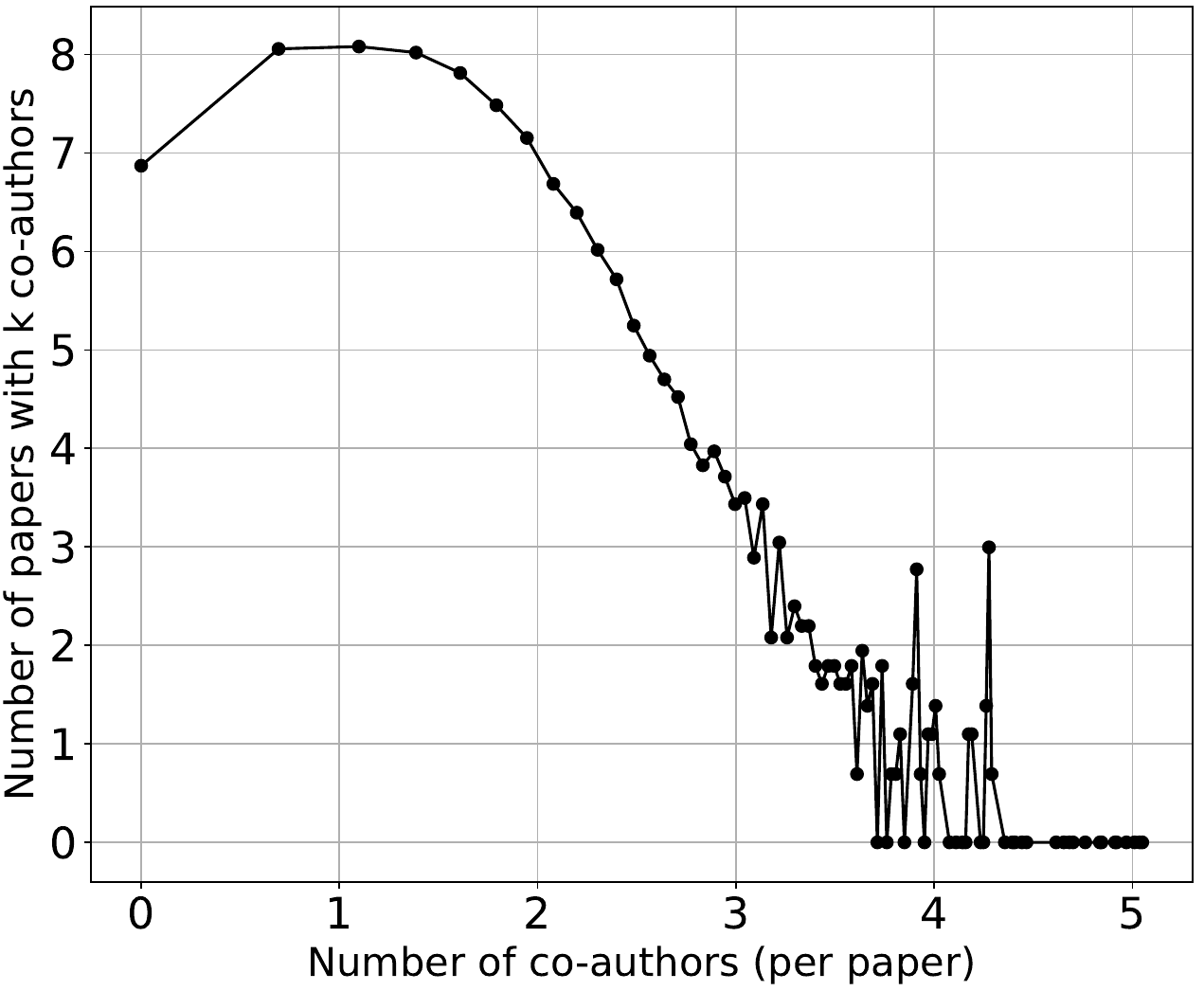}
\caption{With hyperauthors}
\end{subfigure}%
\caption{Log-Normal distribution of the number of authors of a given paper}
\label{fig:papers_dist_log}
\vspace{-0.7cm}
\end{figure}

Figures \ref{fig:coauthor_dist_log} and \ref{fig:papers_dist_log} show the distribution (log-normal) of the number of co-authors of an author's egonetwork and a paper's egonetwork, respectively. The inclusion of hyperauthored papers notably impacts the right-tail of the distribution where a number of authors had a large number of co-authors. As the result, the slope of the right-tail in the (b) network is less steep compared to the (a) network. 
The impact is also visible in the paper egonetwork distribution, with more oscillations along the right-tail. Altogether, this shows that hyperauthorship is the best descriptor of co-authorship network degree distribution due to the high variability of co-authorship counts when hyperauthored papers are included. 
%This shows that the inclusion of hyperauthored papers significantly adds noise to the distribution of co-authorship counts per paper. 

We observe notable differences in average centrality measures when hyperauthored papers are included, as shown in Table \ref{tab:network_descriptives}. Both average degree and closeness centrality increased, 89\% and 3\% respectively. Average eigenvector centrality and betweenness centrality decreased significantly, 76\% and 100\% respectively. It's important to note that while the change in average centrality values seems small, the magnitude of the change is notable given that the values are averages over a large number of observations. 

We further examined how centrality measures are impacted by the presence of hyperauthored papers when different weighting functions are used in the calculation. Table \ref{tab:network_centrality} shows the average centrality measures based on full counting (i.e. ``weighted'') and two partial counting methods, Newman's and Jaccard's functions. We also include the unweighted measures to compare with the weighted counterparts. The percentage change is reported to show difference in measures when hyperauthored papers are included, and the optimal weighting function is one that can minimize this percentage difference. We find that for degree centrality, Newman weighting is most effective in minimizing the difference in measures across the two networks. For betweenness centrality, full counting is preferred as there is no difference in betweenness centrality across the two networks when this weighting function is used. For closeness centrality, Jaccard weighting along with the unweighted measure are preferred with the least difference in closeness centrality when hyperauthored papers are included. For eigenvector centrality, full counting method is preferred as there is no change in centrality in the presence of hyperauthorship. 

Altogether, the findings suggest that including hyperauthored papers distorts micro-level and egocentric measures while maintaining relative stability for the network as a whole. This conclusion is based on the observation that while there is little change in whole-network structure (despite growth in network size), there is significant change in the average centrality measures at the micro-level, indicating that the inclusion of hyperauthored papers can greatly affect the position and influence of individual authors within the network.

%For degreee centrality, the highest increase is found when the network is Jaccard-weighted (+152\%), followed by the unweighted network (+89\%). When the network is Newman-weighted, degree centrality slightly decreased (-5\%). Jaccard-weighted network also yields a notable increase in eigenvector centrality (+100\%) when hyperauthored papers are included. On the other hand, eigenvector centrality values of unweighted and Newman-weighted networks both decreased (-76\% and -33\% respectively). There is no change in eigenvector centrality when the network is weighted via full counting. There is also no change in the betweenness centrality in the weighted network with full counting. The unweighted network, on the other hand, has a significant increase in betweenness centrality value (+100\%). We observe a decrease in betweenness centrality for both Newman-weighted and Jaccard-weighted networks (-33\% and -50\%, respectively). 
%Closeness centrality values increase across all weighting functions, with the magnitude of increase being highest when the network is Newman-weighted. 

\begin{table}[htb]
\centering
\caption{Average centrality measures with various weighting functions for networks \underline{without} and networks \underline{with} hyperauthorship}
\label{tab:network_centrality}
\resizebox{\textwidth}{!}{%
\begin{tabular}{lrrr} 
\toprule
Network measures & \textbf{W/o}~hyperauthors & \textbf{With} hyperauthors & \% change \\ \hline
Avg. Degree Centrality (unweighted) & 13.10 & 24.78 & +89.18\% \\
Avg. Degree Centrality (weighted) & 19.23 & 34.72 & +80.56\%\\
Avg. Degree Centrality (Newman weighted) & 3.05 & 2.90 & -5.08\% \\
Avg. Degree Centrality (Jaccard weighted) & 4.61 & 11.67 & +152.97\% \\ \hline

Avg. Betweenness Centrality (unweighted) & 0.0001 & 0.000 & -100\%\\
Avg. Betweenness Centrality (weighted) & 0.0001 & 0.0001 & 0\% \\
Avg. Betweenness Centrality (Newman weighted) & 0.0003 & 0.0002 & -33.33\%\\
Avg. Betweenness Centrality (Jaccard weighted) & 0.0002 & 0.0001 & -50\%\\  \hline

Avg. Closeness Centrality (unweighted) & 0.23 & 0.24 & +3.90\% \\
Avg. Closeness Centrality (weighted) & 0.20 & 0.21 & +5.97\% \\
Avg. Closeness Centrality (Newman weighted) & 1.52 & 2.30 & +51.85\%\\
Avg. Closeness Centrality (Jaccard weighted) & 21.61 & 22.27 & +3.06\% \\ \hline

Avg. Eigenvector Centrality (unweighted) & 0.003 & 0.0007 & -76.67\% \\
Avg. Eigenvector Centrality (weighted) & 0.0003 & 0.0003 & 0\% \\
Avg. Eigenvector Centrality (Newman weighted) & 0.0003 & 0.0002 & -33.33\%\\
Avg. Eigenvector Centrality (Jaccard weighted) & 0.0002 & 0.0004 & +100\%\\

\bottomrule
\end{tabular}
}
\vspace{-0.5cm}
\end{table}

%Changes in degree centralization --> adding 2 papers added 3-4 extra people (degree centralization) 

\subsubsection{Egocentric network case study}

\begin{table}[htb]
\caption{Egocentric networks that were impacted by the inclusion of hyperauthorship}
\centering
\resizebox{0.60\textwidth}{!}{%
\begin{tabular}{|c|c|cc|}
\hline
 &  & \multicolumn{2}{c|}{With hyperauthor} \\ \hline
 &  & \multicolumn{1}{c|}{High centrality} & Low centrality \\ \hline
\multirow{2}{*}{\begin{tabular}[c]{@{}c@{}}Without\\ hyperauthor\end{tabular}} & High centrality & \multicolumn{1}{c|}{Node 67} & Node 135 \\ \cline{2-4} 
 & Low centrality & \multicolumn{1}{c|}{Node 16} & Node 3918 \\ \hline
\end{tabular}
}
\label{tab:egocentric_cases}
\vspace{-0.5cm}
\end{table}

Given our initial findings that hyperauthor papers produce meaningful structural influences for ego-centric measures, we conducted an egocentric case study of a specific set of authors to explore how their positions in the network changed due to the inclusion of hyperauthorship. The authors are selected based on their importance in the network based on degree centrality (selection criteria shown in Table \ref{tab:egocentric_cases}). Degree centrality is a reliable indicator of power and prestige in our network as authors with high degree centrality are more likely to benefit from their immediate co-authors and their respective co-authorship networks and in terms of knowledge and skills \parencites{badar2016research}{li2013co}. High number of connections within the network also suggests that these authors are actively collaborating and contributing to the field. 

\begin{figure}[htbp]
\begin{subfigure}[h]{0.5\linewidth}
\includegraphics[width=\linewidth]{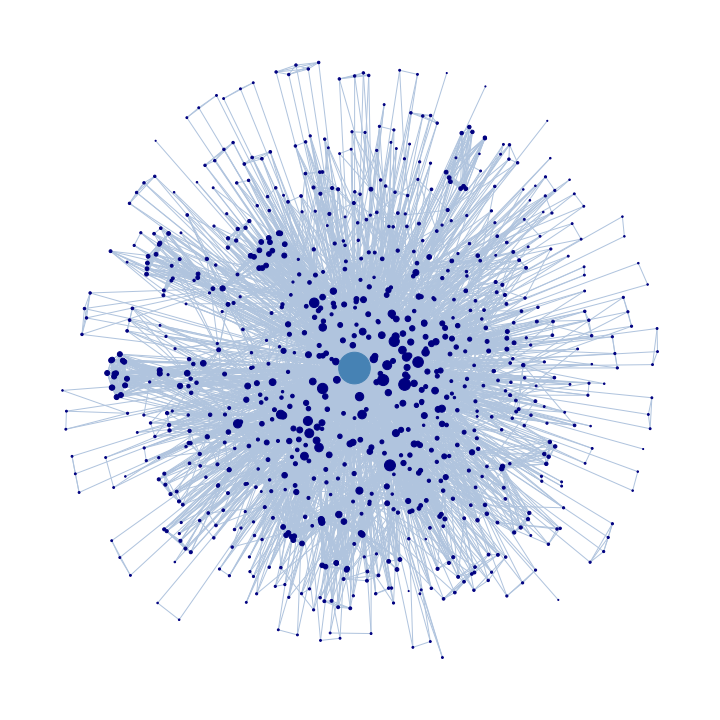}
\caption{Without hyperauthors (weighted)}
\end{subfigure}
\hfill
\begin{subfigure}[h]{0.5\linewidth}
\includegraphics[width=\linewidth]{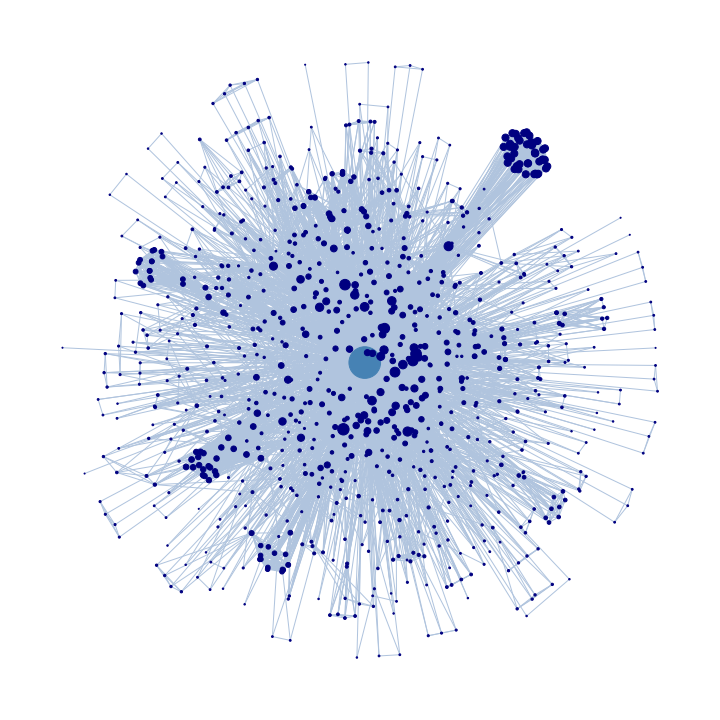}
\caption{With hyperauthors \\ (weighted)}
\end{subfigure}%
\hfill
\begin{subfigure}[h]{0.5\linewidth}
\includegraphics[width=\linewidth]{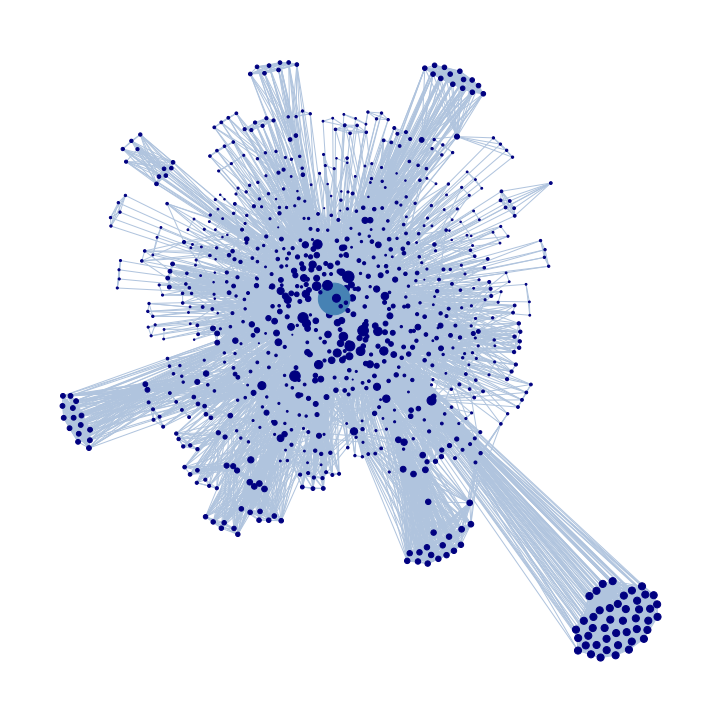}
\caption{With hyperauthors (Newman weighted)}
\end{subfigure}%
\hfill
\begin{subfigure}[h]{0.5\linewidth}
\includegraphics[width=\linewidth]{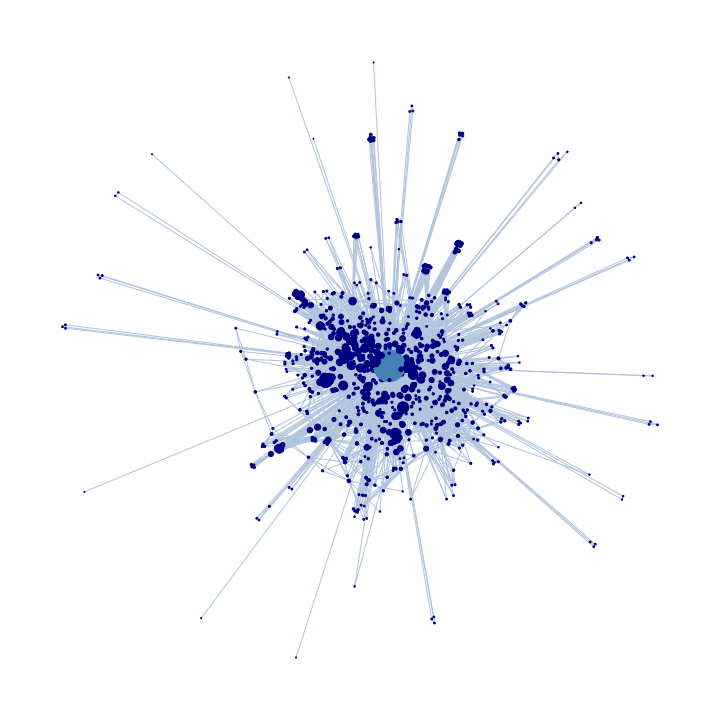}
\caption{With hyperauthors (Jaccard weighted)}
\end{subfigure}%
\caption{Node 67 Ego-networks. Node colors: {\color{RoyalBlue}{Ego}}; {\color{Blue}{Alter}}. Node sized by degree centrality. Without hyperauthor: Nodes=850, Edges=5,425; With hyperauthor: Nodes=916; Edges=6,896}
\label{fig:node67}
\end{figure}

\begin{figure}[htbp]
\begin{subfigure}[h]{0.5\linewidth}
\includegraphics[width=\linewidth]{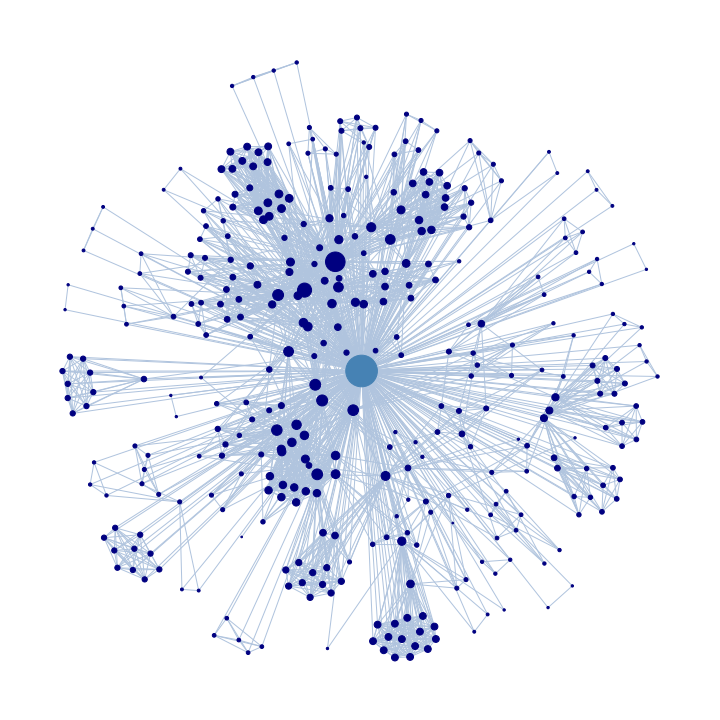}
\caption{Without hyperauthors (weighted)}
\end{subfigure}
\hfill
\begin{subfigure}[h]{0.5\linewidth}
\includegraphics[width=\linewidth]{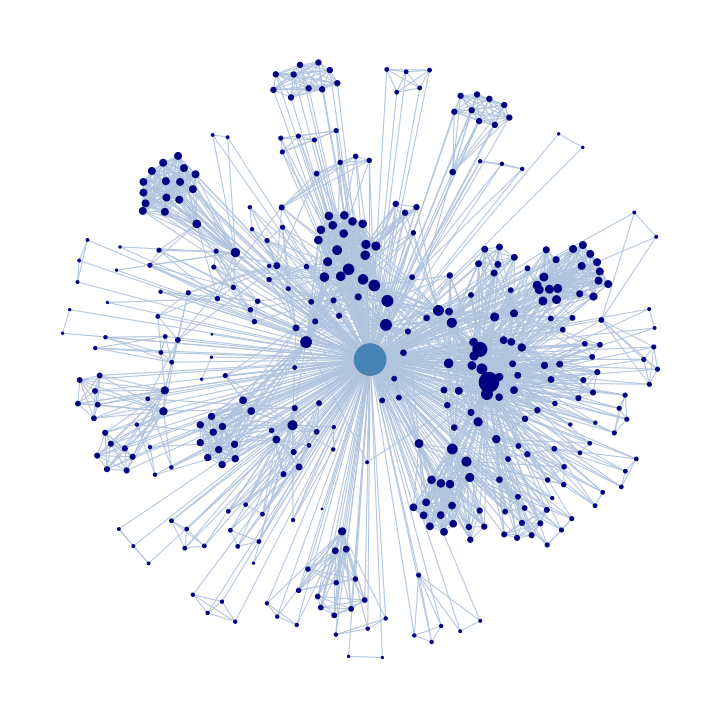}
\caption{With hyperauthors \\ (weighted)}
\end{subfigure}%
\hfill
\begin{subfigure}[h]{0.5\linewidth}
\includegraphics[width=\linewidth]{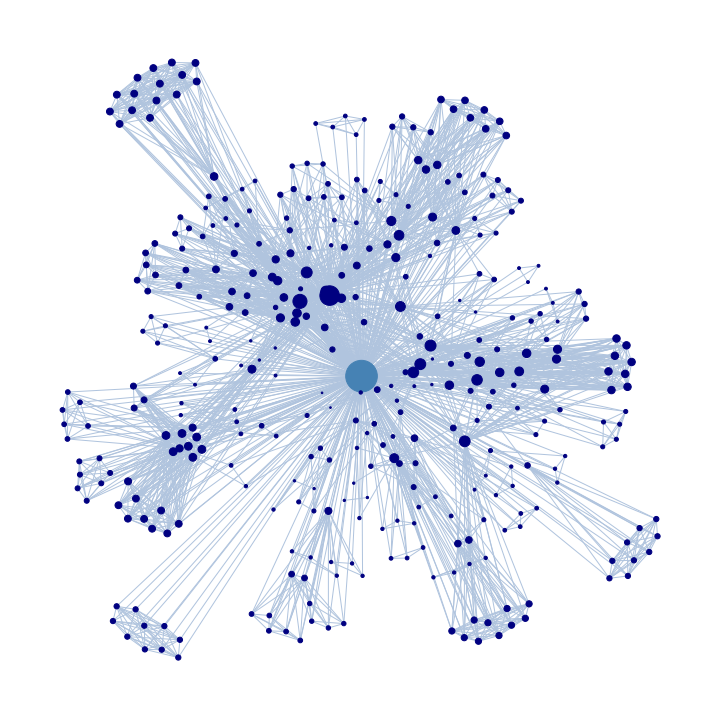}
\caption{With hyperauthors (Newman reweighted)}
\end{subfigure}%
\hfill
\begin{subfigure}[h]{0.5\linewidth}
\includegraphics[width=\linewidth]{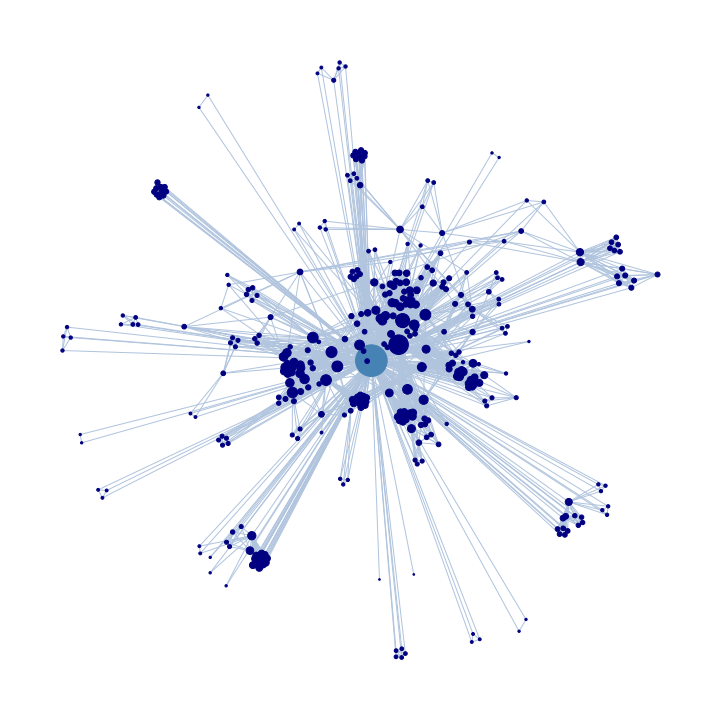}
\caption{With hyperauthors (Jaccard reweighted)}
\end{subfigure}%
\caption{Node 135 Ego-networks. Node colors: {\color{RoyalBlue}{Ego}}; {\color{Blue}{Alter}}. Node sized by degree centrality. Node sized by degree centrality. Without hyperauthor: Nodes=364, Edges=2,297; With hyperauthor: Nodes=364 ; Edges=2,298}
\label{fig:node135}
\end{figure}

\begin{figure}[htbp]
\begin{subfigure}[h]{0.5\linewidth}
\includegraphics[width=\linewidth]{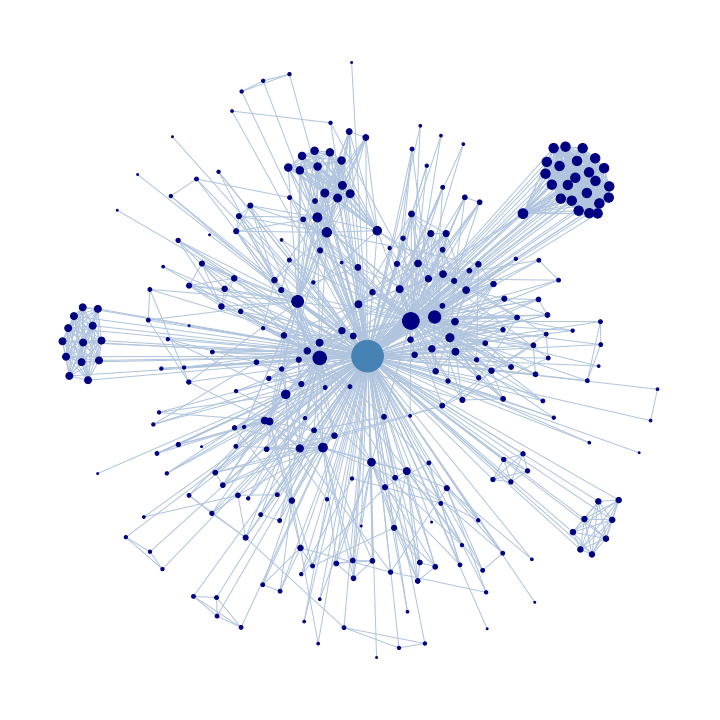}
\caption{Without hyperauthors (weighted)}
\end{subfigure}
\hfill
\begin{subfigure}[h]{0.5\linewidth}
\includegraphics[width=\linewidth]{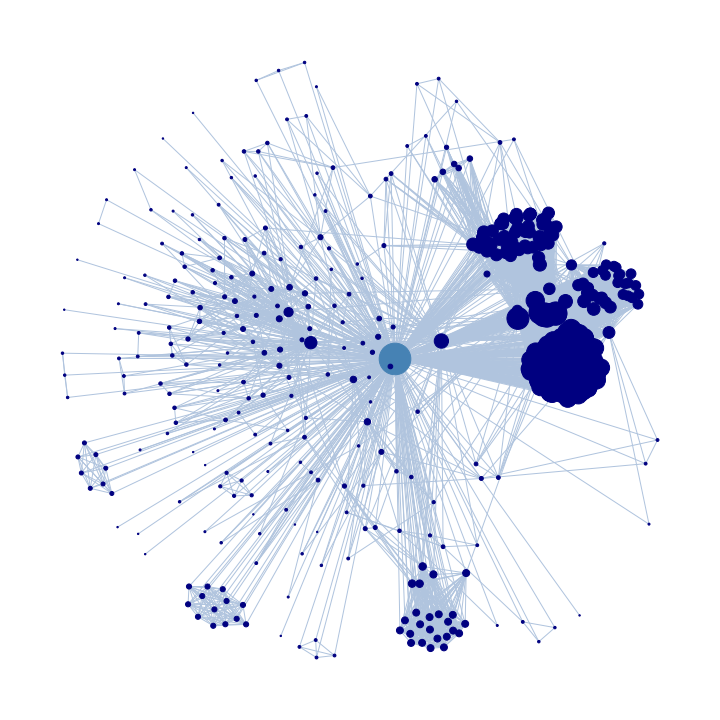}
\caption{With hyperauthors \\ (weighted)}
\end{subfigure}%
\hfill
\begin{subfigure}[h]{0.5\linewidth}
\includegraphics[width=\linewidth]{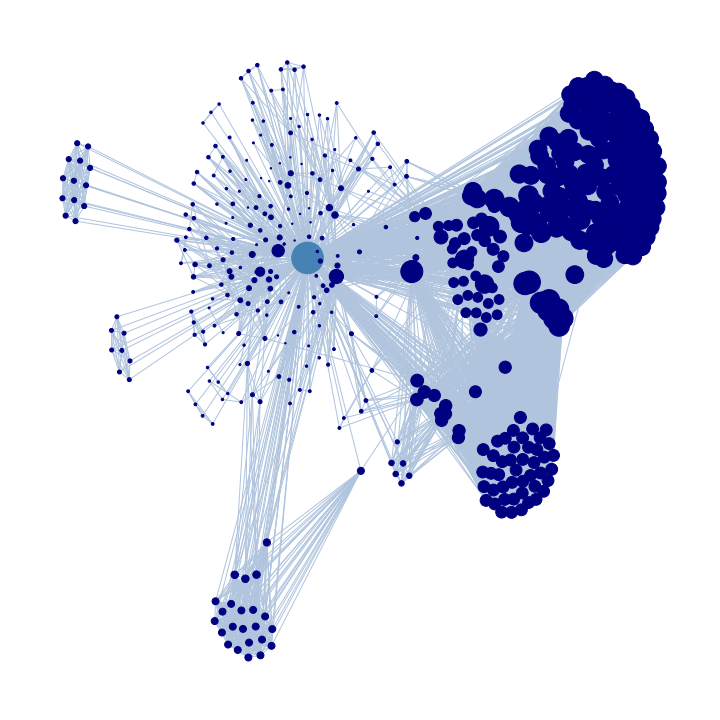}
\caption{With hyperauthors (Newman weighted)}
\end{subfigure}%
\hfill
\begin{subfigure}[h]{0.5\linewidth}
\includegraphics[width=\linewidth]{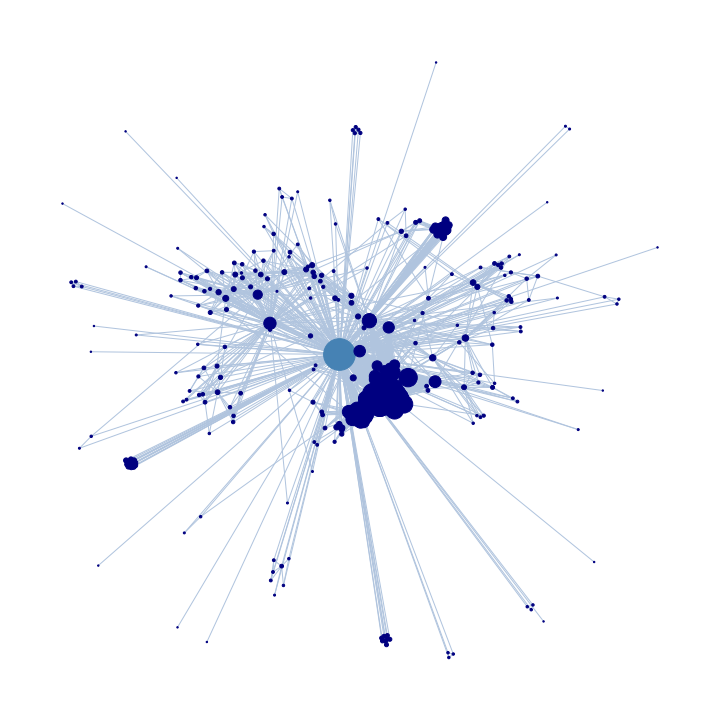}
\caption{With hyperauthors (Jaccard weighted)}
\end{subfigure}%
\caption{Node 16 Egonetworks. Node colors: {\color{RoyalBlue}{Ego}}; {\color{Blue}{Alter}}. Node sized by degree centrality. Node sized by degree centrality. Without hyperauthor: Nodes=276, Edges=1,343; With hyperauthor: Nodes=502 ; Edges=17,284}
\label{fig:node16}
\end{figure}

\begin{figure}[htbp]
\begin{subfigure}[h]{0.5\linewidth}
\includegraphics[width=\linewidth]{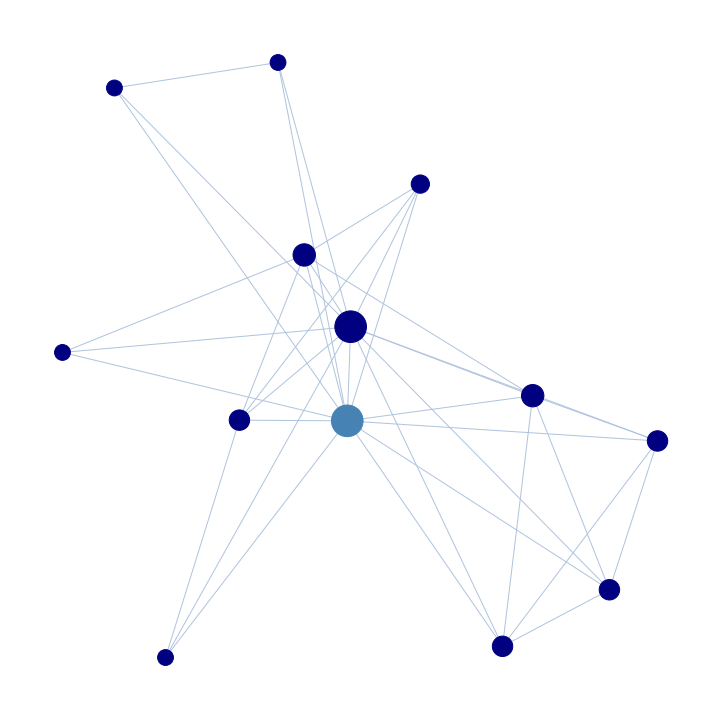}
\caption{Without hyperauthors (weighted)}
\end{subfigure}
\hfill
\begin{subfigure}[h]{0.5\linewidth}
\includegraphics[width=\linewidth]{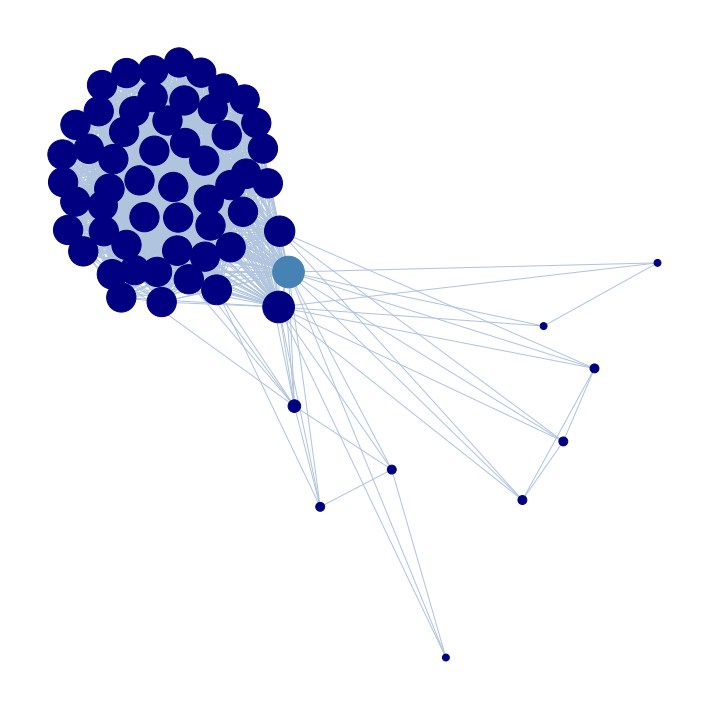}
\caption{With hyperauthors \\ (weighted)}
\end{subfigure}%
\hfill
\begin{subfigure}[h]{0.5\linewidth}
\includegraphics[width=\linewidth]{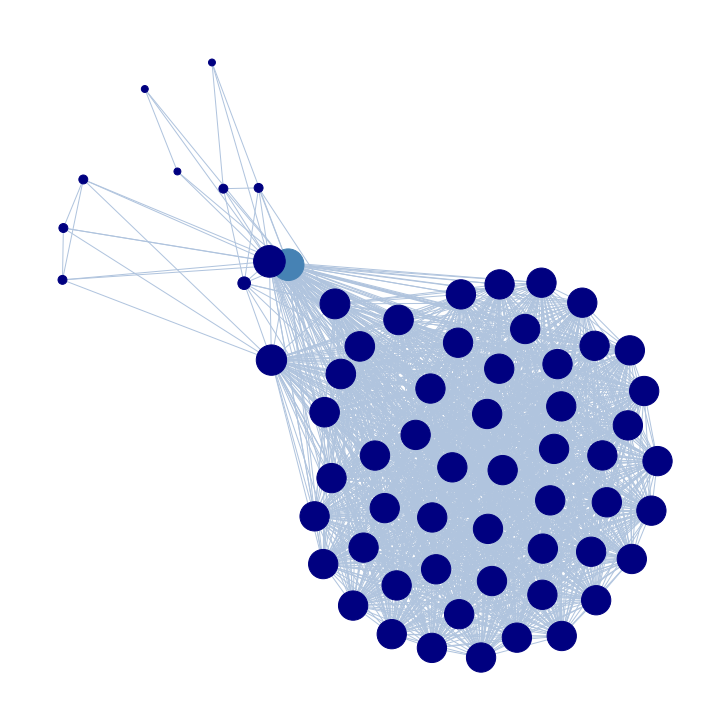}
\caption{With hyperauthors (Newman weighted)}
\end{subfigure}%
\hfill
\begin{subfigure}[h]{0.5\linewidth}
\includegraphics[width=\linewidth]{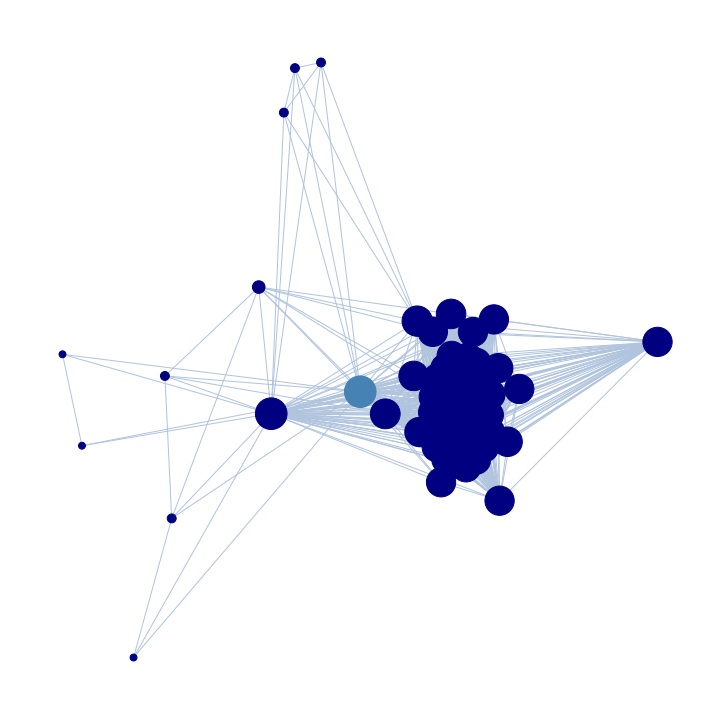}
\caption{With hyperauthors (Jaccard weighted)}
\end{subfigure}%
\caption{Node 3918 Ego-networks. Node colors: {\color{RoyalBlue}{Ego}}; {\color{Blue}{Alter}}. Node sized by degree centrality. Node sized by degree centrality. Without hyperauthor: Nodes=13, Edges=36; With hyperauthor: Nodes=64; Edges=1,521}
\label{fig:node3918}
\end{figure}

Node 67 (egonetworks in Figure \ref{fig:node67}) fits with our selection criteria of high centrality for both networks without and with hyperauthorship as they have high centrality in both networks (rank one and rank three, respectively). In particular, this author produced 747 papers in total, two of which were hyperauthored (one with 27 co-authors, and one with 49 co-authors). Their egonetwork consisted of 850 unique co-authors and 5,425 edges between them when hyperauthorship was excluded. When hyperauthorship is included, their network grew to 916 unique co-authors and 6,896 edges between them. As shown in (a) and (b) of Figure \ref{fig:node67}, the egonetwork size increased slightly with the presence of a large cluster of nodes resulting for hyperauthorship. 

Node 135 (egonetworks in Figure \ref{fig:node135}) provides a case where including hyperauthorship negatively impacted the ego's centrality in the co-authorship network. This node was ranked 28th in degree centrality when excluding hyperauthored works, but dropped to 80th rank when hyperauthorship was included. This is because the author was not involved in any hyperauthored papers. While they published 142 papers (range of co-authors: 0-21) and were central in the network overall, other authors who benefited from hyperauthorship surpassed this author's centrality. As a result, the author's network size did not change, with 364 co-authors and 2,298 edges in the network with hyperauthorship, and 2,297 edges in the network without hyperauthorship. An additional edge appeared in the network with hyperauthorship, representing two authors who published another hyperauthored paper that Node 135 was not a part of. 

Node 16 (egonetworks in Figure \ref{fig:node16}) exemplifies how hyperauthorship can benefit an author's position in the network. The author had 394 papers that were not hyperauthored (range of co-authors: 0-24), and three papers that were hyperauthored, with 49, 70, and 155 co-authors respectively. In the network without hyperauthorship, they had 276 co-authors and 1,343 edges between them. Including their three additional papers in the network with hyperauthorship grew their network to 502 co-authors and 17,284 edges. As a result, this author rose from 52nd rank in degree centrality to 28th rank when hyperauthored works are included. 

Node 3918 (egonetworks in Figure \ref{fig:node3918}) exemplifies a contingency when hyperauthorship may not influence the overall centrality of an author in the network. This author produced six papers without hyperauthored works (range of co-authors: 2-5). Their egonetwork contained 13 unique co-authors and 36 edges between them. When their one hyperauthored paper (with 54 co-authors) was added to the network, their network size grew to 64 unique co-authors and 1,521 edges between them. Even though Node 3918's ranking in terms of degree centrality compared to other authors has remained relatively unchanged with the inclusion of hyperauthorship, their egonetwork grew significantly when a hyperauthored paper was included. 

\subsubsection{Impact of weighting functions}
Having demonstrated the influence of hyperauthored works on standard metrics of centrality, we next examined the impact of weighting functions on the four egonetworks' centrality measures and determined whether certain approaches to weighting centrality assessments offers an optimal method to curtail the inflation effects associated with hyperauthorship. We find that weighting by full and fractional counting significantly changes the four average centrality values of the egonetworks in our case study. 

As shown in Tables \ref{tab:egocentric_centrality_1}-\ref{tab:egocentric_centrality_2} in the Appendix, choices of weighting functions matter to all measures of centrality. In the case of node 67, weighting by full counting yields the highest average degree centrality for networks both without and with hyperauthors. On the other hand, weighting by fractional counting (Newman and Jaccard) brings the average degree centrality down significantly, and even lower than the unweighted degree measures for both networks. Specifically, centrality measures with Jaccard weighting are often the lowest compared to measures from other weighting functions, with the exception of average closeness centrality. The notably high average closeness centrality (173 in the network without hyperauthorship and 180 in the network with hyperauthorship) suggests that there are many authors that receive high scores because their immediate neighbors are well-connected. This effect is magnified in the closeness centrality measure when Jaccard is used as edge weight. For nodes 135, 16, and 3918, we observe similar effects of Jaccard weighting on closeness centrality, where the average closeness measures are notably inflated compared to two other weighting functions. 

We also examined which weighting function(s) are effective in minimizing the percentage change between the network without hyperauthorship and the network with hyperauthorship. This allowed us to determine the optimal weighting function to mitigate the inflated effects of hyperauthorship. We focus on node 16 and node 3918 (results in Table \ref{tab:egocentric_centrality_2} in this analysis because the effects of hyperauthorship were most profound to their egonetworks). In node 16, Newman weighting was most effective as the average degree centrality actually decreased (-23\%) when hyperauthorship is included. For betweenness centrality, unweighted measure was preferred as it best minimizes the percentage change between the two networks. For closeness centrality, unweighted measure yields lowest percentage change while the Newman weighted measure yields the highest percentage change. For eigenvector centrality, Newman weighting was the only function that yields a decrease in eigenvector (-44\%), while other weighting functions are increased due to hyperauthorship. 

We observe similar patterns of results in node 3918, with the exception of betweenness centrality and eigenvector centrality. Jaccard weighting function is effective for betweenness centrality where the percentage decrease is most minimized with this weighting function (-79\%). Eigenvector centrality with Jaccard weighting is also preferred as the impact of hyperauthorship is minimized (-36\%) compared to other weighting functions. Newman weighting function was the most effective in minimizing the inflated effects of hyperauthorship for degree centrality (-52\%), but least effective for closeness centrality (+559\%).

\subsection{Discussion and Conclusion}
Our structural analysis and egocentric analysis of co-authorship networks revealed notable effects that hyperauthorship has on certain centrality measures. First, including even a small number of hyperauthored papers created noise in the overall distribution of number of authors of a given paper (shown in Figure \ref{fig:papers_dist_log}) as well as in the number of co-authors (shown in Figure \ref{fig:coauthor_dist_log}). Secondly, hyperauthorship inflated the average degree and closeness measures, and deflated the average eigenvector and betweenness centrality measures. Structural measures including network size, giant component size, density, and average path length, were inflated as well. Interestingly, average clustering remained unchanged, indicating that while the network grew about twice in size when hyperauthored papers are included, the network is not more connected in terms of local neighborhoods.

Our approach for establishing a hyperauthorship threshold yielded an appropriate cut-off point of 25 authors in our dataset. This threshold value is similar to the threshold of 20 authors as set in \cite{fegley2013has} and \cite{morris2007manifestation}'s studies. On the other hand, this threshold is substantially smaller than the values of 100 authors and 200 authors determined in \cite{cronin2001hyperauthorship} and \cite{milojevic2010modes}'s studies, respectively. We suspect that our cut-off values are different from those in the literature due to differences in the (1) bibliographic database the data are collected from, and (2) the research fields that comprise the datasets. Our data were collected from an internal API that retrieves bibliographic information from Scopus, while others have used PubMed \parencite{fegley2013has}, Web of Science \parencite{morris2007manifestation}, or Thomson Reuters \parencite{milojevic2010modes}. As exemplified in \cite{glanzel2004does}'s study, co-authorship dynamics differ significantly across different research fields. Our data contain publications specific to the field of genomics, while other studies focus on nanotechnology \parencite{milojevic2010modes}, library and information science \parencites{cronin2001hyperauthorship}{morris2007manifestation}, and biomedicine \parencite{fegley2013has}. The observed differences also suggest that determining the hyperauthor cut-off point may depend on the distribution of the dataset. Therefore, a generalizable approach like ours for handling hyperauthorship data would be beneficial, as it could be applied across different disciplines and sources of bibliometric data.

We also examined whether weighting functions can mitigate the impacts of hyperauthorship on centrality network metrics. We compared four different weighting scenarios (i.e. unweighted, weighted based on full counting, weighted based on Newman method, and weighted based on Jaccard method) for the entire network and four egocentric networks. The impact of weighting functions at the whole network-level was slightly different than the impact at the egocentric network-level, and varied based on the centrality measure. For degree centrality, Newman weighting is preferred at both the whole network and egocentric network levels. On the other hand, eigenvector centrality with weighting based on full counting can best curtail the effects of hyperauthorship for both network levels. For betweenness centrality, weighting by full counting is preferred at the whole network level, whereas fractional counting (Newman and Jaccard) is preferred for egocentric networks. For closeness centrality, all weighting methods inflated the measure, and thus no weighting is preferred. Overall, the choices in weighting methods have observable impact on most centrality measures (except closeness centrality) and their resilience to the inflated effects of hyperauthorship. 

Our study contributes to research in bibliometrics and scientific collaborations in specific ways. Our network-analytic approach demonstrates the significant structural impacts that even a small proportion of hyperauthored papers produced for multiple levels of a network, from the egocentric level to the whole network. In particular, we find that degree centrality and closeness centrality are overestimated when hyperauthored papers are included, whereas betweenness and eigenvector centrality are notably underestimated. Thus, our work should be taken as a cautionary message for scholars who are interested in using co-authorship networks to study collaboration. We encourage analysts and readers to think carefully about the nature of the relationships they are studying before deciding whether to include hyperauthored works in their analyses. Our analysis offers several options for how to handle hyperauthored works in analyses.

First, if hyperauthored works are unnecessary for analysis (e.g. because they represent such a small proportion of a dataset), or are semantically distinct from the object of study (e.g. if the analyst wishes to use co-authorship as a proxy for close collaborative relationships), we recommend considering removing them from the dataset. Our paper offers a generalized and mathematically grounded approach for doing so.
Second, if hyperauthored works are important for analysis, we encourage analysts to be mindful when interpreting network statistics that are likely to be inflated by these works. We propose multiple weighting functions to mitigate the impact of hyperauthorship, and find that weighting based on full counting and Newman-based fractional counting are preferred. 

Taken together, our findings are directly relevant to researchers who want to use bibliometrics and network measures to make inferences about scientific collaboration. The removal of hyperauthorship is recommended especially before construction and analysis of co-authorship networks in order to avoid misrepresentation of network density, degree distribution, and centrality ranking of co-authors. Despite the relatively small number of hyperauthored papers in our dataset, their impact on the co-authorship network structure, which serves as a proxy for understanding collaboration, is significant. This effect reflects the changing nature of guidelines and norms that determine who qualifies as an author in a scientific publication \parencite{cronin2001hyperauthorship}. As a result, researchers should be cautious about using co-authorship as a proxy for scientific collaboration, as the nature and extent of each author's contribution in a collaboration can vary widely.

\section{Acknowledgments}
Left blank for review. 

\section{Conflict of Interest Declaration}
The authors declare no conflict of interests. 

\printbibliography

\appendix
\section{{\color{white} appendix}}

\begin{table}[H]
\centering
\resizebox{0.78\textwidth}{!}{%
\begin{tblr}{
  column{3} = {r},
  column{4} = {r},
  column{5} = {r},
  cell{2}{1} = {r=8}{},
  cell{10}{1} = {r=8}{},
  cell{18}{1} = {r=8}{},
  cell{26}{1} = {r=8}{},
  hline{1,34} = {-}{0.08em},
  hline{2} = {2-5}{},
}
 & Network measures & \textbf{W/o~hyperauthors} & \textbf{With hyperauthors} & \% change\\
 
{Node\\67} & Avg. Degree Centrality (unweighted) & 12.750 & 15.040 & +18\%\\
 & Avg. Degree Centrality (weighted) & 23.048 & 24.852 & +8\%\\
 & Avg. Degree Centrality (Newman weighted) & 4.701 & 4.483 & +5\%\\
 & Avg. Degree Centrality (Jaccard weighted) & 2.749 & 4.463 & +62\%\\
 & Avg. Betweenness Centrality (unweighted) & 0.001 & 0.001 & 0\% \\
 & Avg. Betweenness Centrality (weighted) & 0.002 & 0.002 & 0\% \\
 & Avg. Betweenness Centrality (Newman weighted) & 0.002 & 0.002 & 0\% \\
 & Avg. Betweenness Centrality (Jaccard weighted) & 0.001 & 0.001 & 0\%\\ 
 & Avg. Closeness Centrality (unweighted) & 0.504 & 0.504 & 0\% \\
 & Avg. Closeness Centrality (weighted) & 0.380 & 0.386 & +1.6\% \\
 & Avg. Closeness Centrality (Newman weighted) & 2.254 & 2.513 & 11\% \\
 & Avg. Closeness Centrality (Jaccard weighted) & 173.359 & 180.879 & +4\% \\
 & Avg. Eigenvector Centrality (unweighted) & 0.025 & 0.017 & -32\% \\
 & Avg. Eigenvector Centrality (weighted)  & 0.013 & 0.013 & 0\% \\
 & Avg. Eigenvector Centrality (Newman weighted) & 0.011 & 0.011 & 0\% \\
 & Avg. Eigenvector Centrality (Jaccard weighted) & 0.005 & 0.007 & +40\% \\
\hline
 
{Node\\135} & Avg. Degree Centrality (unweighted) & 12.586 & 12.592 & +0.1\%\\
 & Avg. Degree Centrality (weighted) & 17.485 & 17.463 & -0.1\%\\
 & Avg. Degree Centrality (Newman weighted) & 2.535 & 2.535 & 0\%\\
 & Avg. Degree Centrality (Jaccard weighted) & 4.491 & 4.421 & -1.6\%\\
 & Avg. Betweenness Centrality (unweighted) & 0.003 & 0.003 & 0\%\\
 & Avg. Betweenness Centrality (weighted) & 0.004 & 0.004 & 0\%\\
 & Avg. Betweenness Centrality (Newman weighted) & 0.004 & 0.004 & 0\%\\
 & Avg. Betweenness Centrality (Jaccard weighted) & 0.003 & 0.003 & 0\%\\ 
  & Avg. Closeness Centrality (unweighted) & 0.510 & 0.510 & 0\% \\
 & Avg. Closeness Centrality (weighted) & 0.411 & 0.411 & 0\% \\
 & Avg. Closeness Centrality (Newman weighted) & 2.928 & 2.946 & +0.6\% \\
 & Avg. Closeness Centrality (Jaccard weighted) & 47.353 & 47.373 & 0\% \\
 & Avg. Eigenvector Centrality (unweighted) & 0.040 & 0.040 & 0\% \\
 & Avg. Eigenvector Centrality (weighted) & 0.025 & 0.025 & 0\% \\
 & Avg. Eigenvector Centrality (Newman weighted) & 0.022 & 0.022 & 0\% \\
 & Avg. Eigenvector Centrality (Jaccard weighted) & 0.011 & 0.011 & 0\% \\
\end{tblr}
}
\caption{Average centrality measures with different weighting functions for egonetworks (set 1)}
\label{tab:egocentric_centrality_1}
\end{table}

\begin{table}[H]
\centering
\resizebox{0.78\textwidth}{!}{%
\begin{tblr}{
  column{3} = {r},
  column{4} = {r},
  column{5} = {r},
  cell{2}{1} = {r=8}{},
  cell{10}{1} = {r=8}{},
  cell{18}{1} = {r=8}{},
  cell{26}{1} = {r=8}{},
  hline{1,34} = {-}{0.08em},
  hline{2} = {2-5}{},
}
 & Network measures & \textbf{W/o~hyperauthors} & \textbf{With hyperauthors} & \% change\\
 
{Node\\16} & Avg. Degree Centrality (unweighted) & 9.732 & 68.860 & +607\%\\
 & Avg. Degree Centrality (weighted) & 16.812 & 79.992 & +376\%\\
 & Avg. Degree Centrality (Newman weighted) & 4.023 & 3.094 & -23\%\\
 & Avg. Degree Centrality (Jaccard weighted) & 3.664 & 28.793 & +686\%\\
 & Avg. Betweenness Centrality (unweighted) & 0.004 & 0.001 & -75\%\\
 & Avg. Betweenness Centrality (weighted) & 0.005 & 0.002 & -60\% \\
 & Avg. Betweenness Centrality (Newman weighted) & 0.006 & 0.003 & -50\%\\
 & Avg. Betweenness Centrality (Jaccard weighted) & 0.004 & 0.002 & -50\% \\ 
 & Avg. Closeness Centrality (unweighted) & 0.510 & 0.541 & +6\% \\
 & Avg. Closeness Centrality (weighted)  & 0.395 & 0.459 & +16\% \\
 & Avg. Closeness Centrality (Newman weighted) & 1.739 & 4.049 & +133\% \\
 & Avg. Closeness Centrality (Jaccard weighted) & 116.563 & 144.772 & +24\% \\
 & Avg. Eigenvector Centrality (unweighted) & 0.038 & 0.027 & +29\%\\
 & Avg. Eigenvector Centrality (weighted) & 0.021 & 0.029 & +38\% \\
 & Avg. Eigenvector Centrality (Newman weighted) & 0.018 & 0.010 & -44\%  \\
 & Avg. Eigenvector Centrality (Jaccard weighted) & 0.017 & 0.023 & +35\% \\
\hline
{Node\\3918} & Avg. Degree Centrality (unweighted) & 5.538 & 47.531 & +758\%\\
 & Avg. Degree Centrality (weighted) & 11.692 & 49.531 & +324\%\\
 & Avg. Degree Centrality (Newman weighted) & 3.789 & 1.836 & -52\% \\
 & Avg. Degree Centrality (Jaccard weighted) & 1.132 & 29.627 &  +2517\% \\
 & Avg. Betweenness Centrality (unweighted) & 0.049 & 0.004 & -92\% \\
 & Avg. Betweenness Centrality (weighted) & 0.067 & 0.005 & -93\%\\
 & Avg. Betweenness Centrality (Newman weighted) & 0.069 & 0.005 & -93\% \\
 & Avg. Betweenness Centrality (Jaccard weighted) & 0.077 & 0.016 & -79\% \\
  & Avg. Closeness Centrality (unweighted) & 0.673 & 0.831 & +23\% \\
 & Avg. Closeness Centrality (weighted) & 0.583 & 0.810 & +39\% \\
 & Avg. Closeness Centrality (Newman weighted) & 2.247 & 14.805 & +559\% \\
 & Avg. Closeness Centrality (Jaccard weighted) & 25.167 & 53.828 & +114\% \\
 & Avg. Eigenvector Centrality (unweighted) & 0.262 & 0.117 & +55\% \\
 & Avg. Eigenvector Centrality (weighted) & 0.193 & 0.118 & -39\%  \\
 & Avg. Eigenvector Centrality (Newman weighted) & 0.193 & 0.046 & -76\% \\
 & Avg. Eigenvector Centrality (Jaccard weighted) & 0.168 & 0.107 & -36\% \\
\end{tblr}
}
\caption{Average centrality measures with different weighting functions for egonetworks (set 2)}
\label{tab:egocentric_centrality_2}
\end{table}

\end{document}